\documentclass[aps,prd,twocolumn,superscriptaddress,nofootinbib]{revtex4-2}

\usepackage[utf8]{inputenc}
\usepackage[T1]{fontenc}
\usepackage{amsmath,amssymb,amsfonts,mathtools}
\usepackage{bm}
\usepackage{graphicx}
\usepackage{xcolor}
\usepackage{hyperref}
\hypersetup{colorlinks=true,linkcolor=blue,citecolor=blue,urlcolor=blue}

\newcommand{\apjl}{Astrophys.\ J.\ Lett.}
\newcommand{\aap}{Astron.\ Astrophys.}
\newcommand{\mnras}{Mon.\ Not.\ R.\ Astron.\ Soc.}
\newcommand{\aapr}{Astron.\ Astrophys.\ Rev.}
\newcommand{\apss}{Astrophys.\ Space Sci.}
\newcommand{\jcap}{J.\ Cosmol.\ Astropart.\ Phys.}

\begin{document}

\title{Fingerprints of Individual Supermassive Black Hole Binaries in Pulsar Timing Arrays}

\author{Chiara M. F. Mingarelli}
\affiliation{Department of Physics, Yale University, New Haven, 06520, CT, USA}
\affiliation{Center for Computational Astrophysics, Flatiron Institute, 162 5th Avenue, New York, NY, 10010, USA}
\email{chiara.mingarelli@yale.edu}
\author{Bjorn Larsen}
\affiliation{Department of Physics, Yale University, New Haven, 06520, CT, USA}
\author{Ellis Eisenberg}
\affiliation{Department of Astronomy, Yale University, New Haven, 06520, CT, USA}
\author{Qinyuan Zheng}
\affiliation{Department of Physics, Yale University, New Haven, 06520, CT, USA}
\author{Forrest Hutchison}
\affiliation{Department of Physics, Yale University, New Haven, 06520, CT, USA}

\date{July 2, 2026}

\begin{abstract}
\noindent With evidence for a nanoHertz gravitational-wave background now established by Pulsar Timing Arrays, the search focuses on identifying individual supermassive black hole binaries. We show that these binaries produce a distinct spatial correlation pattern across the array, acting as a deterministic analogue to the stochastic Hellings \& Downs curve. We derive a closed analytic expression for this single-source overlap reduction function, $\Upsilon_{ab}$, factorizing the signal into a source-dependent amplitude and a purely geometric fingerprint. Using simulated datasets, we demonstrate that this fingerprint breaks the degeneracy between an individual binary and a stochastic background. Including these cross-correlations yields Bayes factors of $1611$ favoring the continuous-wave model over a Hellings \& Downs correlated background model and $159$ favoring the continuous-wave model over an uncorrelated red-noise model.
Furthermore, these new cross-correlations improve sky localization by a factor of $11\times$ over an uncorrelated search. Finally, while coherent matched filtering offers higher theoretical sensitivity, we argue that a cross-correlation-based search for individual binaries provides a robust alternative that hedges against the possibility of overfitting to noise fluctuations by focusing on the evidence for the correlations.
Indeed, the geometric fingerprints we present here  show that spatial correlations can also be used to identify the first nanoHertz gravitational-wave sources.
\end{abstract}

\maketitle
\section{Introduction}
\label{sec:introduction}
Evidence of the nanoHertz gravitational-wave background (GWB) by Pulsar Timing Arrays (PTAs) marks a major milestone in low-frequency astrophysics~\cite{NG15_gwb, Antoniadis2023, reardon_search_2023, Xu2023, Miles2025}. This signal, consistent with the Hellings \& Downs (HD) spatial correlation curve~\cite{hd83, Allen2023}, is widely believed to be the stochastic superposition of inspiraling supermassive black hole binaries (SMBHBs) distributed throughout the universe~\cite{SBS2019, Ming2019, Mingarelli2025, taylor2025}. With the background established, the observational frontier has now shifted to the identification of the individual, deterministic SMBHBs that comprise it~\cite{ACC2024, ng15-cw, ng15-targeted}.

Detecting these individual continuous wave (CW) sources presents a different challenge than the background. Standard PTA searches for the GWB rely heavily on spatial correlations between pulsars to distinguish gravitational waves (GWs) from intrinsic pulsar noise. In contrast, existing searches for individual binaries typically treat the signal as a deterministic waveform in each pulsar individually, as in the frequentist $\mathcal{F}$-statistic~\cite{Ellis_2012, Babak}. This statistic uses the pulsar antenna patterns to weight the signal, however it tests for phase coherence in the time domain and is therefore distinct from the correlation-based frameworks used for stochastic searches. This creates a conceptual methods gap: we view the background as a spatially correlated phenomenon, yet treat individual binaries as isolated time series linked only by geometric coefficients.

We show that this distinction is artificial: a single bright binary is simply the deterministic limit of a stochastic background. Just as an isotropic population of binaries produces the universal HD correlation curve, we show that a single binary must produce a specific, deterministic spatial correlation pattern across the array. While previous work has used numerical response functions or harmonic decompositions to model this anisotropy~\cite{m13, ani2020, ani2021, cornishsesana2013, GairEtAl2014, Schult2025}, a representation of the precise geometric structure of a single source's correlation pattern in a closed analytic form has remained implicit.

In this work, we derive this missing analytic link. We show that a single SMBHB produces a robust, direction-dependent spatial coherence pattern -- a geometric ``fingerprint'' -- that uniquely characterizes the source. We introduce the single-source overlap reduction function (ORF), $\Upsilon_{ab}(\hat{\bm{\Omega}},\iota,\psi)$, which factorizes the CW signal into a pulsar-independent amplitude and this purely geometric term. This fingerprint acts as the deterministic analogue to the stochastic HD curve. By explicitly modeling these cross-correlations, we show that one can break the degeneracy between a single bright binary and a stochastic background, identifying the specific geometry of the source even when the signal power is dominated by auto-correlations.

The remainder of this paper is organized as follows. In Section~\ref{sec:fingerprints}, we derive the timing response of a single circular SMBHB in a convenient computational frame, obtaining the closed-form expression for the geometric fingerprint $\Upsilon_{ab}$. While spectral or numerical equivalents of this response have been utilized in recent studies (e.g., \cite{Schult2025, cornishsesana2013}), our derivation provides a compact time-domain analytic expression useful for geometric intuition. In Section~\ref{sec:otherwork}, we connect our results to recent studies on PTA anisotropy and hotspots, showing how these features arise from the superposition of single-source fingerprints. In Section~\ref{sec:injections}, we validate the analytic model using simulated datasets, demonstrating how the cross-correlated signature of a CW can be distinguished from uncorrelated noise or an HD-correlated GWB. 
Finally, in Section~\ref{sec:discussion}, we discuss the implications of this method for the current era of PTA detections, comparing the robustness of these geometric fingerprints against fully coherent approaches. A series of appendices collects the supporting calculations. Appendix~\ref{appendix:beam} gives the explicit antenna-pattern functions and the rotation of an arbitrary array into the computational frame. Appendix~\ref{app:averaged_upsilon} derives the inclination and polarization angle averaged single-source ORF. Appendix~\ref{sec:pulsar-terms-CW} discusses when pulsar terms aid or hinder CW analyses. Appendix~\ref{sec:AppC-circ} treats circular polarization in general relativity (GR).

\section{Correlation functions for CWs}
\label{sec:fingerprints}
Our focus in this section is to isolate the geometric structure of the Earth term response.
We focus on the Earth term to isolate the robust, time-invariant geometric fingerprint, which remains detectable even when distance uncertainties decohere the pulsar term. As discussed in Appendix~\ref{sec:pulsar-terms-CW}, the pulsar term introduces additional phase evolution and
pulsar-dependent modulation but does not change the angular shape of the spatial correlation.
The full Earth–plus–pulsar response is reinstated in Sec.~\ref{sec:injections} when we test these results in simulated
PTA datasets with injections and recoveries.

We begin with the standard expression for the timing response of pulsar $a$ to a plane GW with 
propagation direction $\hat{\bm{\Omega}}$~\citep{Detweiler1979, anholm2009}. The fractional frequency shift induced by the GW is
\begin{equation}
z_a(t) = \frac{1}{2}
\frac{\hat{p}_a^i \hat{p}_a^j}{1 + \hat{\bm{\Omega}}\!\cdot\!\hat{\bm{p}}_a}
\Delta h_{ij}(t,\hat{\bm{\Omega}}) ,
\label{eq:redshift-general}
\end{equation}
where $\hat{\bm{p}}_a$ is the unit vector from the Earth to pulsar $a$, and $\Delta h_{ij}$ is the 
difference in the GW metric perturbation between the Earth and pulsar terms.
The corresponding timing residual is
\begin{equation}
s_a(t) = \int^t \! dt' \, z_a(t') .
\label{eq:residual-general}
\end{equation}
Latin indices run over spatial components and repeated indices are summed.

With the pulsar term neglected, we set $\Delta h_{ij}(t,\hat{\bm{\Omega}}) \to h_{ij}(t,\hat{\bm{\Omega}})$,
the metric perturbation evaluated at the Earth, which we decompose into plus and cross polarizations as
\begin{equation}
h_{ij}(t,\hat{\bm{\Omega}}) =
\sum_{A=+,\times} h_A(t)\,e^A_{ij}(\hat{\bm{\Omega}}) ,
\label{eq:metric-decomposition}
\end{equation}
where $e^A_{ij}$ are the usual plus and cross polarization tensors.
Substituting Eq.~\eqref{eq:metric-decomposition} into Eq.~\eqref{eq:redshift-general}, and
integrating in time, the timing residual for pulsar $a$ can be written as
\begin{equation}
s_a(t) = \sum_{A} F_a^A(\hat{\bm{\Omega}})\,s_A(t) ,
\label{eq:residual-decomposition}
\end{equation}
where the antenna pattern functions are
\begin{equation}
F_a^A(\hat{\bm{\Omega}}) =
\frac{1}{2}
\frac{\hat{p}_a^i \hat{p}_a^j}
{1 + \hat{\bm{\Omega}}\!\cdot\!\hat{\bm{p}}_a}
e^A_{ij}(\hat{\bm{\Omega}}) ,
\label{eq:antenna-pattern}
\end{equation}
and the polarization-dependent residuals are
\begin{equation}
s_A(t) = \int^t \! dt' \, h_A(t') .
\label{eq:polarization-residuals}
\end{equation}
Equations~\eqref{eq:residual-decomposition}–\eqref{eq:polarization-residuals} make explicit the
separation between the purely geometric response encoded in $F_a^A(\hat{\bm{\Omega}})$ and the
source-dependent time evolution contained in $h_A(t)$. Eq. \eqref{eq:antenna-pattern} is sometimes defined with an additional degree of freedom allowed for rotations of the polarization tensors by angle $\psi$ (or alternatively, rotation of observation frame). We will proceed at first assuming $\psi = 0$ for clarity of exposition and generalize our results to arbitrary $\psi$ later on. We note specifically that certain results where $\psi = 0$ will also hold for cases where $\psi$ is marginalized over, as discussed in Appendix~\ref{sec:AppC-circ}.

\subsection{The CW Signal Model}
\label{sec:signalmodel}
We consider a circular, non-evolving binary with GW angular frequency $\omega_0 = 2\pi f_0$ and 
chirp mass $\mathcal{M}_c$ at luminosity distance $D_L$. In a polarization basis aligned with 
the binary orbital angular momentum, the plus and cross strains at the Earth are~\citep{SesanaVecchio2010_CW, corbincornish2010}
\begin{align}
h_+(t) &= h_0 (1+\cos^2\iota)\cos(\omega_0 t + \phi_0) ,\\
h_\times(t) &= -2 h_0 \cos\iota \sin(\omega_0 t + \phi_0) ,
\end{align}
where $\iota$ is the inclination angle, $\phi_0$ is an initial phase, and
\begin{equation}
h_0 = \frac{2 \mathcal{M}_c^{5/3}}{D_L} (\pi f_0)^{2/3}
\end{equation}
in geometric units~\citep{peters_gravitational_1963, Ellis_2012, cornishsesana2013}.  The orbital evolution is slow over a PTA baseline, justifying this non-evolving treatment of the frequency. Indeed, negligible frequency evolution over the observation baseline requires
$\dot f_0\,T_{\rm obs} \ll f_0$. 
To show this, we use the quadrupole prediction for a circular binary~\citep{peters_gravitational_1963},
\begin{equation}
\dot f_0 = \frac{96}{5}\pi^{8/3}\mathcal{M}_c^{5/3} f_0^{11/3},
\end{equation}
for which a fiducial SMBHB with $f_0 = 6\,\mathrm{nHz}$, $\mathcal{M}_c \simeq 10^9 M_\odot$, and $T_{\rm obs} = 16\,{\rm yr}$ gives
\[
\frac{\dot f_0\,T_{\rm obs}}{f_0} \sim 3\times 10^{-5} \ll 1.
\]
Frequency evolution is therefore negligible for the sources considered here. We also assume
GR with only $+$ and $\times$ tensor polarizations, though alternative polarizations are also possible and are explored in~\citet{Zheng2026}.

Integrating once in time produces the polarization-dependent timing residuals
\begin{align}
s_+(t) &= \frac{h_0}{\omega_0} (1+\cos^2\iota)\sin(\omega_0 t + \phi_0) ,\\
s_\times(t) &= \frac{2 h_0}{\omega_0} \cos\iota \cos(\omega_0 t + \phi_0) ,
\end{align}
where we have dropped constant terms that are absorbed by the timing model, and we define the amplitude as
\begin{equation}
A_{\rm CW} \equiv \frac{h_0}{2\pi f_0}
= \frac{h_0}{\omega_0}
= \frac{1}{2\pi f_0} \frac{2 \mathcal{M}_c^{5/3}}{D_L} (\pi f_0)^{2/3} .
\label{eq:A-CW-def}
\end{equation}

Using Eq.~\eqref{eq:residual-decomposition}, the residual in pulsar $a$ can then be written as
\begin{equation}
s_a(t) = A_{\rm CW}
\left[
\alpha_a \sin(\omega_0 t + \phi_0)
+ \beta_a \cos(\omega_0 t + \phi_0)
\right] ,
\label{eq:sa-alpha-beta}
\end{equation}
where
\begin{align}
\alpha_a &= (1+\cos^2\iota)\,F_a^+(\hat{\bm{\Omega}}) ,\\
\beta_a  &= 2\cos\iota\,F_a^\times(\hat{\bm{\Omega}}) .
\end{align}
The redshift $z_a(t)$ is proportional to $h_A(t)$, while the observable residual $s_a(t)$ differs 
only by this time integration and by the geometric coefficients carried by the antenna patterns.

It is useful to rewrite Eq.~\eqref{eq:sa-alpha-beta} as
\begin{equation}
s_a(t) = A_{\rm CW} A_a
\sin(\omega_0 t + \phi_0 + \delta_a) ,
\label{eq:sa-A-delta}
\end{equation}
with
\begin{align}
A_a &= \sqrt{\alpha_a^2 + \beta_a^2} ,\\
\delta_a &= \arctan\!\left(\frac{\beta_a}{\alpha_a}\right) \, .
\end{align}
The amplitude $A_a$ and phase offset $\delta_a$ are functions of the source 
orientation and the geometric factors $F_a^A$. Eq.~\eqref{eq:sa-alpha-beta} is equivalent to Eq.~21 in \citet{cornishsesana2013} and Eq.~28 in \citet{Schult2025}, here evaluated in the computational frame to facilitate the derivation of $\Upsilon_{ab}$.

The cross correlation between pulsars $a$ and $b$ is
\begin{equation}
C_{ab} \equiv \langle s_a(t)\,s_b(t) \rangle ,
\end{equation}
where the brackets denote a time average. Using Eq.~\eqref{eq:sa-A-delta}, we can write
\begin{equation}
s_a(t)=A_{\rm CW} A_a \sin(\omega_0 t+\alpha),
s_b(t)=A_{\rm CW} A_b \sin(\omega_0 t+\beta),
\end{equation}
with $\alpha=\phi_0+\delta_a$ and $\beta=\phi_0+\delta_b$. The finite-time average over 
an observation span $T$,
\begin{equation}
\langle X(t) \rangle_T \equiv \frac{1}{T}\int_0^T dt\,X(t),
\end{equation}
gives
\begin{align}
\langle s_a(t)s_b(t)\rangle_T
&= \frac{A_{\rm CW}^2 A_a A_b}{T}\nonumber\\
&\quad\times\int_0^T dt\,\sin(\omega_0 t+\alpha)\sin(\omega_0 t+\beta).
\end{align}
Using $\sin x\sin y=\tfrac12[\cos(x-y)-\cos(x+y)]$ and integrating, we obtain
\begin{align}
&\langle s_a(t)s_b(t)\rangle_T
=\frac{A_{\rm CW}^2 A_a A_b}{2}\cos(\delta_a-\delta_b)\nonumber\\
&\; -\frac{A_{\rm CW}^2 A_a A_b}{4\omega_0 T}
\Big[\sin(2\omega_0 T+\alpha+\beta)-\sin(\alpha+\beta)\Big].
\label{eq:Cab-finiteT}
\end{align}
The second term is a boundary term suppressed by $1/(\omega_0 T)$. In terms of the GW 
period $P\equiv 2\pi/\omega_0$ and the number of observed cycles 
$N_{\rm cyc}\equiv T/P=\omega_0 T/(2\pi)$, the fractional correction obeys
\begin{equation}
\frac{|\Delta C_{ab}|}{A_{\rm CW}^2 A_a A_b/2}
\;\lesssim\; \frac{1}{2\pi N_{\rm cyc}}.
\end{equation}
For $N_{\rm cyc}\gtrsim 3$ the boundary term is at the few-percent level or smaller, which 
is the relevant PTA regime. In the limit $T\to\infty$ it vanishes, and the cycle-averaged 
correlation takes the simple form
\begin{equation}
C_{ab}
= \frac{A_{\rm CW}^2}{2}\,A_a A_b \cos(\delta_a-\delta_b).
\label{eq:Cab-factorized-clean}
\end{equation}
We can therefore write the single-source ORF
\begin{equation}
\Gamma_{ab}(\hat{\bm{\Omega}},\iota)
\equiv A_a A_b \cos(\delta_a-\delta_b),
\label{eq:upsilon-def-clean}
\end{equation}
which depends only on the geometry of the source and the two pulsars $a$ and $b$. 
Here, the inclination $\iota$ is implicitly carried by $A_a$ and $\delta_a$. To be clear, Eq.~\eqref{eq:upsilon-def-clean} recovers the result that the cross correlation of two sine waves scales as the cosine of the phase difference.

Using Eq.~\eqref{eq:upsilon-def-clean} and the amplitude–phase form, the overlap reduction 
function simplifies to
\begin{equation}
\Gamma_{ab}(\hat{\bm{\Omega}},\iota)
= A_a A_b \cos(\delta_a-\delta_b)
= \alpha_a\alpha_b + \beta_a\beta_b.
\end{equation}
Substituting the definitions of $\alpha_a$ and $\beta_a$ gives
\begin{equation}
\Gamma_{ab}(\hat{\bm{\Omega}},\iota)
= (1+\cos^2\iota)^2 F_a^+F_b^+ 
+ 4\cos^2\iota\,F_a^\times F_b^\times.
\label{eq:gamma-raw}
\end{equation}
Here the antenna patterns are defined to be aligned with the polarization axes, which is why the $\psi$ dependence is suppressed. To restore $\psi$-dependence, we need to transform the antenna pattern functions by rotating them in the transverse plane,
\begin{align}
    \begin{pmatrix}
        F^+ \\
        F^{\times}
    \end{pmatrix}_{\psi=0} = \begin{pmatrix}
        \cos2\psi & \sin2\psi \\
        -\sin2\psi & \cos2\psi
    \end{pmatrix}\begin{pmatrix}
        F^+ \\
        F^{\times}
    \end{pmatrix}.
\end{align}
Applying to Eq.~\eqref{eq:gamma-raw} yields the $\psi$-dependent cross correlation,
\begin{align}
    2\Gamma_{ab}(\hat{\bm{\Omega}},\iota,\psi) &= \left[1 + 6\cos^2\iota + \cos^4\iota\right](F^+_aF^+_b + F^\times_aF^\times_b) +\nonumber \\
    &\quad \sin^4\iota\cos(4\psi)(F^+_aF^+_b - F^\times_aF^\times_b) +\nonumber \\
    &\quad \sin^4\iota\sin(4\psi)(F^+_aF^\times_b + F^\times_aF^+_b).
    \label{eq:gamma_psi}
\end{align}
The inclination and polarization angle dependence are non-trivially built into the cross correlation, and they determine the correlation pattern together with the system geometry. However, there are many degeneracies between the two parameters. For example, $\psi$-dependence is completely removed for a face-on binary, i.e. when $\sin\iota = 0$, in which case Eq.~\eqref{eq:gamma-raw} and Eq.~\eqref{eq:gamma_psi} become equivalent. Moreover, any case of an edge-on binary ($\cos\iota = 0$) where $\psi=0$ (or $\psi$ is an integer multiple of $n/2$) retains only the plus polarization $h_+$, thus we recover a special case of the form Eq.~\eqref{eq:gamma-raw}. One may also marginalize over $\iota$ and $\psi$ to make the cross correlation purely geometric, which may be useful in applications such as optimal statistics (for details, see Appendix~\ref{app:averaged_upsilon}). As such, the geometric relations given by Eq.~\eqref{eq:gamma-raw} holds for statistically unpolarized point GW sources as well as certain binary orientations corresponding to a linear combination of circular and linear GW polarizations where only the plus component of linear polarization contributes. A SMBHB with arbitrary orientation, corresponding to the most general elliptically polarized GW, requires the full Eq.~\eqref{eq:gamma_psi}.

We show in Appendix~\ref{sec:pulsar-terms-CW} that the pulsar term introduces additional phase
evolution and pulsar dependent modulation, but in the regime where pulsar distances are
poorly known it does not alter the shape of the cross correlation. The geometric
fingerprint $\Gamma_{ab}(\theta,\phi,\iota)$ is therefore fully determined by the Earth term, while
the full Earth plus pulsar response is required only when modeling the detailed phase
structure in PTA datasets.

\subsection{Cross correlation in the computational frame}
\label{sec:upsilon-derivation}

\begin{figure}
    \centering
    \includegraphics[width=\linewidth]{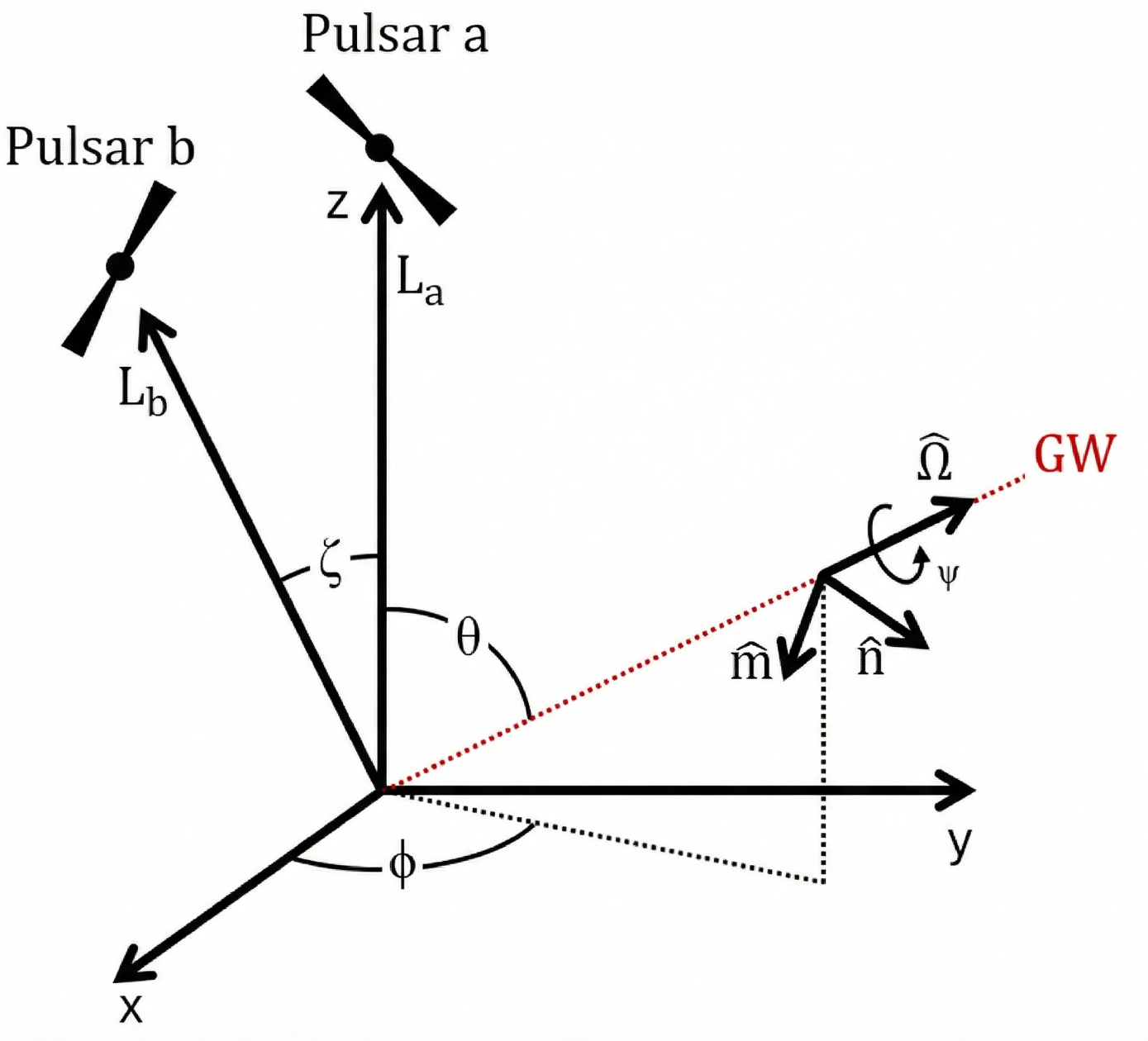}
    \caption{Computational frame used to define the single-source overlap reduction function $\Upsilon_{ab}(\hat{\bm{\Omega}},\iota,\psi)$. Pulsar $a$ lies on the $+\hat{\mathbf{z}}$ axis and pulsar $b$ lies in the $x$–$z$ plane at an angular separation $\zeta$. The GW propagation direction $\hat{\bm{\Omega}}$ is specified by polar angle $\theta$ and azimuthal angle $\phi$ with respect to $+\hat{\mathbf{z}}$. The direction to the GW source is $-\hat{\bm{\Omega}}$. This fixed geometry determines the antenna pattern functions $F_a^{A}$ and $F_b^{A}$ and hence the spatial correlation $\Upsilon_{ab}$ used throughout the paper. Shown also are the GW polarization basis vectors $\hat{\bm{m}}$ and $\hat{\bm{n}}$, which may be subject to a rotation by angle $\psi$.}
    \label{fig:ptaGeometry}
\end{figure}

Unlike the Hellings \& Downs curve, we cannot yet write the ORF, $\Gamma_{ab}$, as a direct function of the angular sky separation between pulsars, $\zeta$, as there is additional amplitude modulation depending on location of each individual pulsar relative to the GW source location. However, it is still possible to express the ORF as a geometric function of $\zeta$ on the level of individual pulsar pairs by rotating each pair into its own \emph{computational frame} with a particular geometry (see Fig.~\ref{fig:ptaGeometry}) \citep{m13, MS14}. In the computational frame, a given pair and the GW source vectors are rotated such that pulsar $a$ lies on the $+\hat{\bm{z}}$ axis and pulsar $b$ 
lies in the $x$–$z$ plane at separation $\zeta$,
\begin{align}
\hat{\bm{p}}_a &= (0,\,0,\,1),\\
\hat{\bm{p}}_b &= (\sin\zeta,\,0,\,\cos\zeta).
\end{align}
To distinguish from the cosmic rest frame, we notate computational frame-defined ORF $\Gamma \to \Upsilon$, although we emphasize the value of the cross correlation is unchanged as it depends only on the relative geometry between each pulsar and the GW source. The GW propagation direction is parameterized by angles $(\theta,\phi)$ with respect to 
$\hat{\bm{z}}$,
\begin{equation}
\hat{\bm{\Omega}}
= (\sin\theta\cos\phi,\;\sin\theta\sin\phi,\;\cos\theta),
\end{equation}
and points from the source toward the observer. Since $\hat{\bm{\Omega}}$ is the propagation direction, $(\theta,\phi)$ point opposite to the GW source, which instead lies along $-\hat{\bm{\Omega}}$. This is the canonical definition of $\hat{\bm{\Omega}}$, however it is opposite to the convention in software such as \texttt{enterprise}~\citep{ENTERPRISE}, where the sky angles point toward the source. 
The inclination angle $\iota$ is defined only with respect to the GW plane of propagation and remains unchanged after the transformation, whereas the polarization angle $\psi$, in principle, must be rotated alongside sky location coordinates.

Full expressions for the antenna pattern functions $F_a^A$ and $F_b^A$ are given in 
Appendix~\ref{appendix:beam}, but for the present derivation we need only two facts:
(i) in this computational frame $F_a^\times=0$, and (ii) both $F_a^A$ and $F_b^A$ are 
trigonometric functions of $(\theta,\phi,\zeta)$.

\begin{figure*}[ht!]
    \centering
    \includegraphics[width=0.49\linewidth]{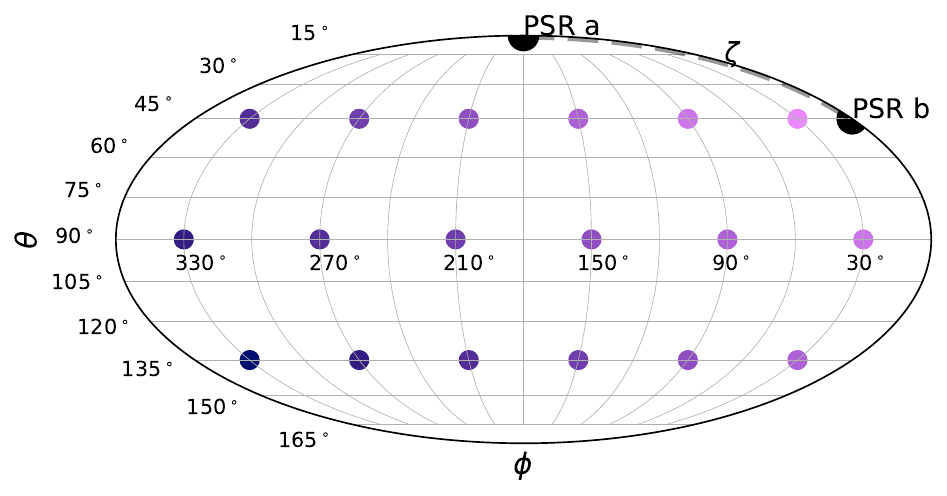}
    \includegraphics[width=0.49\linewidth]{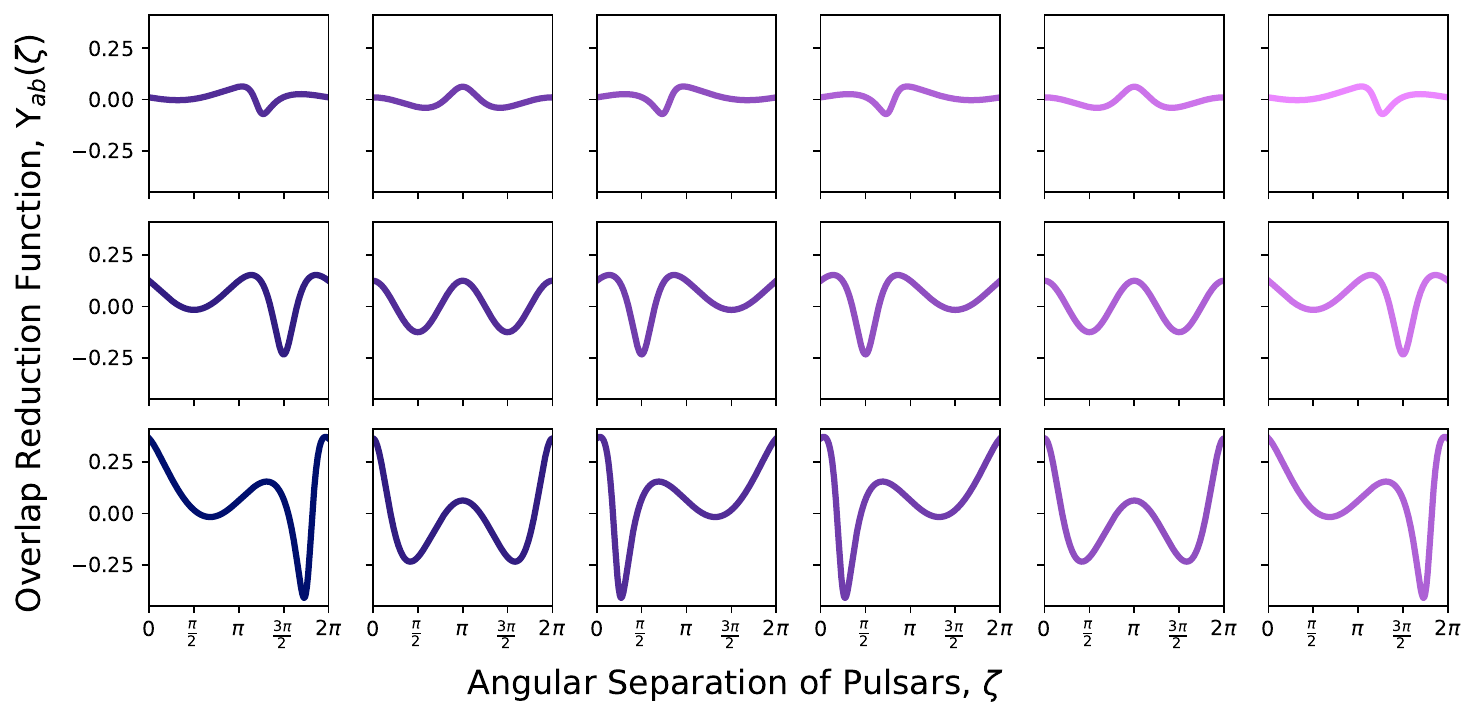}
    \caption{
    Single source correlation fingerprints across the sky. Each colored point on the left marks a GW propagation direction in the computational frame $(\theta,\phi)$, and the curve with the same color on the right shows the corresponding ORF for a single orientation-marginalized GW source, $\Upsilon_{ab}(\hat{\bm{\Omega}})$, Eq.~\eqref{eq:upsilon-final}, for a representative pulsar pair as a function of their angular separation $\zeta$. Pulsar $a$ lies on the $+\hat{\bm{z}}$ axis and pulsar $b$ lies in the $x$-$z$ plane. Unlike the HD curve, these responses are not universal: their shapes depend on source position because different pulsar pairs probe different lobes and nodes of the quadrupolar antenna pattern, and they are periodic, rather than symmetric, about $\pi$ radians due to the innate anisotropy of the single GW source. %
    Particular binary orientations will further distort the cross correlation in a $\psi$ and $\iota$-dependent way, following Eq.~\eqref{eq:upsilon-final-psi}. %
    Each SMBHB therefore imprints a distinct ``fingerprint'' on PTA correlations that encodes its sky position and polarization, which can be used to identify individual binaries and distinguish their signals from, e.g. the GWB.
    }
    \label{fig:upsilon_sky}
\end{figure*}

\begin{figure}[ht!]
\centering
\includegraphics[width=\linewidth]{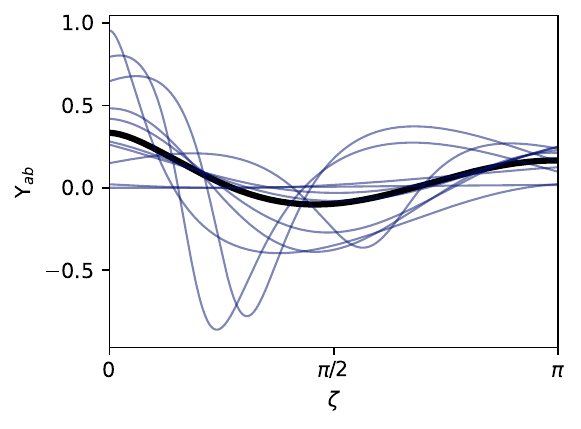}
\caption{
Single source correlation fingerprints (Eq.~\ref{eq:upsilon-final}) corresponding to CWs from ten randomly drawn sky locations, marginalized over $\iota$ and $\psi$. For comparison, the unnormalized Hellings \& Downs curve is shown in black.}
\label{fig:random_draws}
\end{figure}

We will start with the general case of a CW produced by a SMBHB with arbitrary orientation. The observation that $F_a^\times=0$ reduces Eq.~\ref{eq:gamma_psi} to
\begin{align}
    2\Gamma_{ab}(\hat{\bm{\Omega}},\iota,\psi) &= \left[1 + 6\cos^2\iota + \cos^4\iota\right]F^+_aF^+_b +\nonumber \\
    &\quad\quad \sin^4\iota\cos(4\psi)F^+_aF^+_b+\nonumber \\
    &\quad\quad \sin^4\iota\sin(4\psi)F^+_aF^\times_b.
\end{align}
Substituting the explicit antenna patterns from 
Appendix~\ref{appendix:beam} then yields the ORF directly as a function of $\zeta$ alongside sky location and orientation parameters as
\begin{widetext}
\begin{align}
\Upsilon_{ab}(\theta,\phi,\zeta, \iota, \psi) &= (1+6\cos^2\iota+\cos^4\iota+\sin^4\iota\cos4\psi)\frac{(1 - \cos\theta)}{8}\frac{(\cos\phi\cos\theta\sin\zeta -\sin\theta\cos\zeta)^2 - \sin^2\phi\sin^2\zeta}{1 + \cos\phi\sin\theta\sin\zeta + \cos\theta\cos\zeta}\notag\\
&-\sin^4\iota\sin4\psi\frac{(1-\cos\theta)}{4}\frac{\sin\zeta\sin\phi(\cos\theta\sin\zeta\cos\phi-\sin\theta\cos\zeta)}{1+\cos\theta\cos\zeta+\sin\theta\sin\zeta\cos\phi}.
\label{eq:upsilon-final-psi}
\end{align}
\end{widetext}
This expression defines the most general geometric fingerprint of a single circular SMBHB in the computational frame. 
We may also examine the special case where binary orientation is marginalized over, which as previously discussed, also applies for the special cases of a circularly polarized GW, a linearly polarized GW with no $h_\times$ component, or linear combination thereof. As shown in Appendix~\ref{sec:AppC-circ}, we are free to also set $\psi = 0$ in the computational frame for the derivation of this result. Starting from Eq.~\eqref{eq:gamma-raw}, $F_a^\times=0$ simplifies the ORF to
\begin{align}
\Upsilon_{ab}(\hat{\bm{\Omega}},\iota)
&= (1+\cos^2\iota)^2 F_a^+F_b^+.
\end{align}
This shows that the inclination angle in fact introduces only a scaling factor to the ORF, which can be factored into the spectral amplitude component to yield a purely sky-dependent ORF,
\begin{align}
\Upsilon_{ab}(\hat{\bm{\Omega}}) &\propto F_a^+F_b^+.
\end{align}
Despite that we assumed $\psi = 0$ explicitly, this form is equivalent up to a normalization to marginalization over $\psi$ shown in Appendix~\ref{app:averaged_upsilon}, i.e. the ORF for an unpolarized GW point anisotropy, which has been previously derived in the literature \citep{anholm2009, cornishsesana2013, Schult2025}. Substituting the explicit antenna patterns from Appendix~\ref{appendix:beam} and simplifying gives the ORF (up to a constant scaling factor) as a function of $\zeta$ alongside only source sky location parameters,
\begin{widetext}
\begin{equation}
\Upsilon_{ab}(\theta,\phi,\zeta) = \frac{(1 - \cos\theta)}{4}\frac{(\cos\phi\cos\theta\sin\zeta -\sin\theta\cos\zeta)^2 - \sin^2\phi\sin^2\zeta}{1 + \cos\phi\sin\theta\sin\zeta + \cos\theta\cos\zeta}.
\label{eq:upsilon-final}
\end{equation}
\end{widetext}
Eq.~\eqref{eq:upsilon-final} provides a simplified use case when marginalizing over or assuming particular binary orientations, see Figs.~\ref{fig:upsilon_sky}, \ref{fig:random_draws}. These results provide closed-form analytic expressions for the single-source ORF as functions of $\zeta$, complementing the implementations found in previous anisotropic studies~\cite{cornishsesana2013, m13, Schult2025, KumarKamion2024}.

We emphasize that Eqs.~(\ref{eq:upsilon-final-psi}, \ref{eq:upsilon-final}) are derived in the specific computational frame where pulsar $a$ lies on the $z$-axis and pulsar $b$ lies in the $xz$-plane. To apply this to a general array, the source coordinates $(\theta, \phi)$ as well as $\psi$ must be rotated into this specific frame for each pulsar pair, analogous to the frame rotation used for spherical harmonic decompositions in~\citet{m13} and related harmonic analyses~\cite{Nay2024, ng15-harmonic, Allen2024}.
For fixed pulsar separation $\zeta$, inclination $\iota$, and polarization angle $\psi$, the function $\Upsilon_{ab}(\hat{\bm{\Omega}}, \iota, \psi)$ on
the sky $(\theta,\phi)$ encodes the spatial correlation pattern induced by the binary.

Several features are worth noting.
The denominator contains the familiar factor $1 + \hat{\bm{\Omega}}\!\cdot\!\hat{\bm{p}}_b$, which regularizes the response as the GW direction approaches the pulsar line of sight.
The numerator contains products of $\sin\theta$ and $(\cos\theta - 1)$ that enforce a quadrupolar structure, and explicit factors of $\sin\zeta$ that vanish when the two pulsars are coincident. When this occurs, the ORF reduces to the following in the computational frame,
\begin{align}
    \Upsilon_{a=b}(\theta)
    &\propto F_a^+F_a^+ \propto \sin^4\!\left(\frac{\theta}{2}\right),
    \label{eq:Upsilon-auto}
\end{align}
which vanishes when the GW source is opposite the pulsar on the sky ($\theta=0$) and peaks when the source and pulsar are aligned ($\theta=\pi$).

While the normalization of the HD curve is set to yield its interpretation as a cross-correlation coefficient, $\Upsilon_{ab}$ cannot be generically scaled in the same way since the amplitude of the autocorrelation in pulsars $a$ and $b$ depends on their locations relative to the GW source. %

\subsection{Binaries with orbital eccentricity}
\label{sec:eccentricity}
Although this work focuses on SMBHBs in circular orbits, it is useful to clarify how
eccentricity modifies the signal. An eccentric binary does not emit at a single
frequency, but instead radiates at a discrete set of harmonics
$f_n = n f_{\rm orb}$ whose relative amplitudes are set by the Peters and Mathews
decomposition~\cite{peters_gravitational_1963}. Each harmonic behaves like a circular
binary at its corresponding frequency, with its own mixture of $+$ and $\times$
polarizations. The timing response is therefore a sum of circular components rather
than a single monochromatic mode.

Writing the timing residual schematically as
\begin{equation}
s_a(t)
= A_{\rm orb}\sum_{n=1}^{\infty}
A_{a,n}\sin\!\big(n\omega_{\rm orb} t + \delta_{a,n}\big),
\end{equation}
with $\omega_{\rm orb} = 2\pi f_{\rm orb}$ and coefficients $A_{a,n}$ and $\delta_{a,n}$
that encode the geometric response and polarization content of each harmonic, the cross
correlation generalizes to
\begin{equation}
C_{ab}
= \frac{A_{\rm orb}^2}{2}
\sum_{n=1}^{\infty}
A_{a,n}A_{b,n}\cos\!\big(\delta_{a,n}-\delta_{b,n}\big),
\end{equation}
a weighted sum of circular fingerprints. Only in the circular limit, where the
radiation is dominated by the quadrupolar $n=2$ harmonic, does this reduce to the
single-mode expression derived in Sec.~\ref{sec:fingerprints}. While the geometric basis functions do not change, the effective ORF, $\Upsilon_{ab}$, is a weighted sum of these harmonics.

For SMBHBs in the PTA band the $n=2$ harmonic typically carries most of the strain
power unless the eccentricity is very large ($e\gtrsim0.7$). This dominance is made clear
in analytic studies of the harmonic structure
\cite{BarackCutler2004, Sesana2010, TaylorHuertaGair2016} and in astrophysical
population modelling of SMBHBs evolving in stellar or gaseous environments
\cite{OLearyKocsisLoeb2009, KelleyBlechaHernquist2017}. In the regime where the
$n=2$ mode dominates, the circular fingerprint $\Upsilon_{ab}(\hat{\bm{\Omega}})$
accurately describes the leading geometric contribution to the spatial correlation.
Higher harmonics add additional structure to the correlation pattern that our model
does not capture, but they do not change the underlying quadrupolar
polarization content or the basic connection to the HD curve when averaged over source
populations. A full treatment of eccentric sources, including the harmonic mixing in
$\Upsilon_{ab}$, is left to future work.

\subsection{Connection to the HD curve}
\label{sec:connecToHD}
The single-source ORF $\Upsilon_{ab}(\hat{\bm{\Omega}},\iota,\psi)$ describes the
spatial correlation pattern produced by one circular SMBHB at sky location $(\theta,\phi)$.
For an isotropic, unpolarized Gaussian background the sky is filled with many such sources
with random positions and orientations. In that case the relevant quantity is the 
sky-averaged correlation as a function only of pulsar separation $\zeta$,
\begin{equation}
\Gamma^{\rm HD}_{ab}(\zeta)
\propto
\int_{S^2} d\hat{\bm{\Omega}}\int_{-1}^1 d(\cos\iota)\int_0^{\pi/2}d\psi\,
\Upsilon_{ab}(\hat{\bm{\Omega}},\iota,\psi),
\label{eq:sky-average}
\end{equation}
up to an overall normalization. This is the standard
ORF for an isotropic background.
Carrying out the angular integral in Eq.~\eqref{eq:sky-average} reproduces the 
HD curve:
\begin{equation}
\Gamma^{\rm HD}_{ab}(\zeta)
= \frac{3}{2}x_{ab}\ln x_{ab} - \frac{x_{ab}}{4} + \frac{1}{2}(1+\delta_{ab}),
\label{eq:HD-curve}
\end{equation}
where $x_{ab} = (1 - \cos\zeta_{ab})/2$, as derived in detail by \citet{anholm2009} and generalized to anisotropic cases in
\citet{m13} and \citet{GairEtAl2014}. See Appendix~\ref{app:averaged_upsilon} for the results of integration over $\psi$ and $\iota$.

This makes the connection between the deterministic and stochastic regimes explicit.
In the limit where one bright binary dominates the sky the PTA measures
$\Upsilon_{ab}(\hat{\bm{\Omega}},\iota,\psi)$, which depends on the specific sky location and orientation of that source.
As more binaries are added with random orientations and positions, the sky average in
Eq.~\eqref{eq:sky-average} approaches the isotropic HD curve
\cite{cornishsesana2013}, and the isotropy of the sky eliminates the requirement to use a computational frame in order to define a $\zeta$-dependent ORF. The single-source fingerprint $\Upsilon_{ab}(\hat{\bm{\Omega}},\iota,\psi)$ is therefore the
building block whose isotropic superposition yields the familiar HD correlation. We note that the numerical evaluation of the integral produces identical results to the analytic forms.

\subsection{Alternative ORFs}
\label{sec:altORFs}

Alongside the spectral content of various GW sources, the cross-correlation patterns are fundamentally important quantities to distinguish between signals, as we later explore in Section~\ref{sec:injections} (see also \citealt{Ferranti+2025}). While Eqs.~\eqref{eq:gamma_psi} and \eqref{eq:HD-curve} define the ORF for a circular SMBHB and an isotropic GWB respectively, an alternative hypothesis is the $\iota$ and $\psi$-marginalized ORF given by Eq.~\ref{eq:upsilon-final}. This has been previously used elsewhere, e.g., by \citet{Schult2025} as a ``spike-pixel'' model of GW anisotropy, which is (up to a normalization) defined in the cosmic rest frame as
\begin{equation}
    \Gamma^{\rm SP}_{ab}(\hat{\bm{\Omega}}_{\rm gw})
    = \frac{1}{2}\left(1+\delta_{ab}\right)
    \sum_{A=+,\times} F_a^A(\hat{\bm{\Omega}}_{\rm gw}) F_b^A(\hat{\bm{\Omega}}_{\rm gw}),
    \label{eq:spike-pixel}
\end{equation}
where the Kronecker delta accounts for the standard PTA convention for auto– and cross–power spectra in PTA likelihoods. While this form has intrinsically less information than Eq.~\eqref{eq:gamma_psi}, \citet{Schult2025} shows Eq.~\ref{eq:spike-pixel} may still be useful to detect and characterize individual binaries, i.e., an individual binary will manifest predominantly as a point anisotropy.

Additionally, searches for cross-correlations of a GWB typically use a \emph{common uncorrelated red noise} (CURN) model, which uses solely pulsar autocorrelations, $\Gamma_{ab} = \delta_{ab}$, as a conservative null hypothesis. This is because any mismodeled intrinsic pulsar noise is expected to only enter the common spectral model through the autocorrelations. As such, a conservative null hypothesis for CW searches is the purely auto–correlated CW–like spectral model,
\begin{equation}
    \Gamma^{\rm CWdiag}_{ab}(\hat{\bm{\Omega}}_{\rm gw})
    = \delta_{ab}\sum_{A=+,\times}\bigl[F_a^A(\hat{\bm{\Omega}}_{\rm gw})\bigr]^2,
    \label{eq:Gamma-auto}
\end{equation}
equivalent to Eq.~\ref{eq:Upsilon-auto} (note this form does not include the $\psi$-dependence, which introduces pair-dependent amplitude modulations). This diagonal model acts as a catch-all for any type of signal or noise that may spuriously arise as a CW and is the most direct CW analog of the CURN model used as a conservative null hypothesis for the GWB. This null hypothesis puts the onus on the distinctive spatial correlations induced by the CW to carry the statistical weight, rather than temporal coherence alone. Like the CURN model, this diagonal model also simplifies inversion of $\bm{\Phi}$. We explore further use of this model in comparison to the other correlated models in Section~\ref{sec:injections}.

\section{Connection to other studies}
\label{sec:otherwork}
Several recent studies have highlighted features that naturally emerge when only a small number of bright SMBHBs contribute power in the PTA band. \citet{Becsy2022} find spectral distortions, anisotropy, and large variance around the HD curve, while \citet{cornishromano2015} show that PTA data often prefer hybrid models with a few deterministic binaries atop a stochastic background. \citet{Schult2025} contains a derivation of the ORF due to a GW point anisotropy, builds a signal model to search for a binary using this ORF, and applies it in realistic simulated datasets. These searches demonstrate that deterministic binaries produce directional hotspots in frequency-resolved anisotropy searches and that Earth-term-only templates underperform at low frequencies compared to anisotropic background models~\cite{Schult2025, petrov2025}. The per-frequency optimal statistic introduced by \citet{Gersbach+2025_PFOS} and the generalized optimal statistic~\cite{Sardesai2024}, along with its extension to frequentist anisotropy searches in \citet{Gersbach+2025_anisotropy}, likewise reveal frequency-dependent sky power influenced by individual bright sources. A complementary spectral approach is developed by \citet{GundersenCornish2025}, who model CW signals through FFT decompositions of deterministic templates rather than cross correlations. Fast Bayesian techniques for individual binary searches have also been developed~\cite{Becsy2022_QCW, Gardiner+2025}. Together, these frameworks provide valuable and mutually reinforcing perspectives on how bright binaries imprint themselves on PTA data.

Our work provides a compact analytic description of the underlying geometry: each bright binary contributes a well-defined spatial correlation pattern $\Upsilon_{ab}(\hat{\bm{\Omega}},\iota,\psi)$, and the features reported in these papers arise from partial or imperfect averaging over these single-source fingerprints. When many $\Upsilon_{ab}(\hat{\bm{\Omega}},\iota,\psi)$ patterns are summed, the result approaches the HD curve. When only a few dominate, the sky retains the geometric structure of the individual fingerprints we derive here.

Cornish and Sesana~\cite{cornishsesana2013} showed that the standard HD correlation curve can also describe signals from sparse, anisotropic populations of SMBHBs, and even from a single bright binary. Their result is derived by binning pairwise correlations as a function of pulsar separation and averaging over many pulsar pairs. In the limit of a large, isotropically distributed array, this averaging reproduces the sky integrals that define the HD curve, so the \emph{mean} correlation as a function of separation is identical for an isotropic Gaussian background and for a single quadrupolar point source. 

Our calculation asks a different question: for a fixed PTA geometry and a fixed circular binary, what is the exact cross-correlation pattern between individual pulsar pairs? The answer is the geometry-dependent fingerprint $\Upsilon_{ab}(\hat{\bm{\Omega}},\iota,\psi)$ derived here, which in general is not HD-like. Only after averaging $\Upsilon_{ab}(\hat{\bm{\Omega}},\iota,\psi)$ over source directions (or equivalently over pulsar locations at fixed separation) does one recover the HD limit, consistent with \citet{cornishsesana2013} and with the mean-correlation results of \citet{Allen2023}. The anisotropy, scatter, and variance around HD seen in realistic SMBHB simulations~\cite{Becsy2022, cornishromano2015, Schult2025} can therefore be understood as incomplete averaging over these single-source fingerprints rather than a failure of the HD framework itself.

The approaches mentioned above, from stochastic modeling to hybrid deterministic–stochastic frameworks, frequency-domain anisotropy searches, and spectral template methods, each illuminate different aspects of the same underlying physics. Our analytic expression for $\Upsilon_{ab}(\hat{\bm{\Omega}},\iota,\psi)$ provides the geometric foundation that unifies these perspectives and clarifies why they all reveal consistent signatures of bright binaries in the nHz band, whether those signatures are expressed as deviations from the HD curve, directional hotspots, or frequency-dependent anisotropy.

\section{Injection and recovery in a simulated PTA}
\label{sec:injections}
We now shift to demonstrating how interpulsar fingerprints for a single CW may be leveraged in a practical Bayesian analysis, using a cross–correlated spectral model for a single CW in an example simulated dataset. Two primary applications of such a model are: (i) searching for CWs with a small number of parameters prior to a template–based analysis, and (ii) testing CW coherence against a conservative null hypothesis such as a purely auto–correlated CW–like model. The second test will be helpful for claiming robust CW detections in PTA datasets, alongside other proposed coherence tests such as those in \citet{Becsy2025}. Both use cases apply to blind and targeted CW searches.

\begin{table}
    \centering
    \begin{tabular}{ c | c @{}}
        \hline\hline Dataset property & Description \\
        \hline $N_{\rm psr}$ & $100$ \\
        \hline $T_{\rm obs}$ & $16$ yr \\
        \hline $N_{\rm TOA}$ & $300$ \\
        \hline $\sigma_{\rm TOA}$ & $0.1$ $\mu$s \\
        \hline $L_p$ & $1$ kpc \\
        \hline\hline Injected CW parameter & Value \\
        \hline $\theta_{\rm gw}$ & $5\pi/7$ \\
        \hline $\phi_{\rm gw}$ & $5\pi/3$ \\
        \hline $\mathcal{M}_c$ & $1.6\times10^9$ M$_\odot$ \\
        \hline $f_{\rm gw}$ & $6$ nHz \\
        \hline $\Phi_0$ & $0$ \\
        \hline $\psi$ & $\pi/6$ \\
        \hline $\iota$ & $\pi/2$ \\
        \hline $h_0$ & $[1.66, 2.71, 3.71, 5.25] \times 10^{-16}$ \\
        \hline Template S/N & $[3,8,15,30]$ \\
        \hline\hline
    \end{tabular}
    \caption{Simulated dataset and injected CW properties. The injected CW includes pulsar terms with random pulsar phases and the pulsar distances fixed to 1 kpc. We generate 4 copies of the dataset each with different $h_0$ corresponding to the template S/N and a random noise realization.}
    \label{tab:simulation}
\end{table}

We use \href{https://github.com/bencebecsy/pta_replicator}{\texttt{pta\_replicator}} to simulate an idealized array of $N_{\rm psr}=100$ pulsars, placed randomly on the sky following a uniform sky distribution, with equal observation timespans, uniform cadence, identical time of arrival (TOA) uncertainties. We create four copies of this dataset, each with a CW injected using the deterministic template model with parameters listed in Table~\ref{tab:simulation}, as well as a random white noise realization and no additional red noise. The CW properties are identical in each simulation except for the signal-to-noise ratio (S/N) using the injected template model, Eq.~\eqref{eq:residual-decomposition}, and the corresponding strain amplitude $h_0$. For our pilot simulation study, we choose an edge-on binary ($\iota = \pi/2$), corresponding to a purely linearly polarized CW, in order to maximize the impacts of the binary orientation angle $\psi$. While we use a noiseless simulation for proof-of-concept, we note that in real data, intrinsic white noise, red noise, and the GWB itself are primary contaminants for CW searches. The utility of the cross-correlated model lies precisely in its orthogonality to the spatially uncorrelated red noise processes that affect individual pulsars.

The injected CW undergoes about three GW cycles across the dataset, so the finite-window error is at most $\sim 5\%$. This setup is intended as a proof of concept; more realistic datasets are left for future work.
We emphasize that this is an idealized, proof-of-concept simulation rather than a forecast for any specific array. Its $N_{\rm psr}=100$ pulsars and $\sigma_{\rm TOA}=0.1\,\mu{\rm s}$ white-noise level modestly exceed those of current PTAs, while the $T_{\rm obs}=16\,$yr baseline is typical of existing datasets.

To model the CW and noise, we adopt the standard Gaussian PTA likelihood,
\begin{align}
    \mathcal{L}(\bm{\delta t}) &=
    \frac{1}{\sqrt{\det(2\pi\bm{C})}}
    \exp\left(-\frac{1}{2}\bm{\delta t}^{T}\bm{C}^{-1}\bm{\delta t}\right),
\end{align}
with covariance
\begin{equation}
    \bm{C} = \bm{D} + \bm{F}\bm{\Phi}\bm{F}^{T}.
\end{equation}
Here $\bm{\delta t}$ is the vector of timing residuals, $\bm{F}$ is the CW spectral design matrix. This is constructed explicitly in Eq.~\eqref{eq:Fdesign} below, and not to be confused with the antenna pattern functions $F_a^A$ of Eq.~\eqref{eq:antenna-pattern}. The $\bm{D}$ matrix collects the contributions from the marginalized timing model and TOA errors,
\begin{equation}
    \bm{D} = \bm{I}\,\sigma_{\rm TOA}^{2} + \bm{M}\bm{X}\bm{M}^{T},
\end{equation}
whose inverse is cached to carry out faster evaluations~\citep{Johnson+2024}. Here $\bm{M}$ is the timing model design matrix and $\bm{X}$ is the diagonal matrix of large priors for the timing model parameters. For this demonstration we adopt a simple five–parameter timing model (constant offset, spin frequency, spin–down, and two sky-location parameters for each pulsar) and do not include additional red or excess white noise terms. We note that the theoretical template S/N used to set the amplitude of the CW in each dataset is computed using the $\bm{D}$ matrix as
\begin{align}
    \label{eq:S/N}
    {\rm S/N} = \frac{1}{2}h_0^2\bm{s}^T\bm{D}^{-1}\bm{s},
\end{align}
which comes from the log of the signal-to-noise likelihood ratio assuming the noise is unresolved, where $\bm{s}$ is the timing residual vector from the template model with strain amplitude factored out.

The CW spectral design matrix $\bm{F}$ is constructed as a sine–cosine pair for every pulsar. Labeling pulsars by $k=1,\dots,N_{\rm psr}$, the two CW basis functions for pulsar $k$ occupy columns $2k-1$ and $2k$:
\begin{align}
    F_{i,2k-1} &= \cos\bigl(2\pi f_{\rm gw} t_{i,{\rm PSR}_k}\bigr),\label{eq:Fdesign}\\
    F_{i,2k}   &= \sin\bigl(2\pi f_{\rm gw} t_{i,{\rm PSR}_k}\bigr),
\end{align}
with $i$ running over all TOAs in pulsar $k$. This explicit construction allows for sub-Fourier-bin resolution by dynamically updating the design matrix at each likelihood evaluation. For an elliptically-polarized GW sourced by a SMBHB, the CW covariance $\bm{\Phi}$ is modeled as
\begin{equation}
    \Phi^{\rm BHB}_{ab,j} = \Gamma^{\rm BHB}_{ab}(\hat{\bm{\Omega}}_{\rm gw},\iota,\psi)
    \frac{h_0^2}{4\pi^2 f_{\rm gw}^2},
    \label{eq:Phi_matrix_CW}
\end{equation}
where $\Gamma^{\rm BHB}_{ab}$ is the single-source ORF in the spectral representation, defined in Eq.~\eqref{eq:gamma_psi}. This is the cosmic rest frame representation of the fingerprint $\Upsilon_{ab}(\hat{\bm{\Omega}},\iota,\psi)$ derived in Section~\ref{sec:upsilon-derivation}: the two are related by a rotation of the source coordinates into the computational frame (Eq.~\eqref{eq:upsilon-final-psi}). We refer to this model as the ``BHB'' model throughout. In this form $\Gamma^{\rm BHB}_{ab}$ encodes the pulsar–dependent antenna pattern of the CW, while the overall amplitude is set by $h_0^2/(4\pi^2 f_{\rm gw}^2)$, consistent with the time–domain factorization in Eq.~\eqref{eq:Cab-factorized-clean}. Note this model is equivalent to the deterministic template model only in the limiting case of no frequency evolution of the binary, which manifests in the pulsar terms.

To test the benefits of the full cross-correlated model, we also compare the CW model from a polarized point source (Eqs.~\ref{eq:Phi_matrix_CW},~\ref{eq:gamma_psi}) against three alternative hypotheses discussed in Sections~\ref{sec:connecToHD} and \ref{sec:altORFs}. These alternative models each use the same covariance structure and design matrix $\bm{F}$, Eq.~\eqref{eq:Phi_matrix_CW}, but a different ORF. 

The first model is a diagonal-$\bm{\Phi}$ model which includes only the autocorrelations of an unpolarized GW point source (Eq.~\ref{eq:Gamma-auto}), which for CW modeling is analogous to the CURN model for a GWB \citep{CornishSampson2016, Becsy2025}. This represents a more conservative null hypothesis for CW searches than a model with pulsar noise only. 

The second is a Hellings \& Downs correlated model (Eq.~\ref{eq:HD-curve}), representing a spike frequency fluctuation of an isotropic GWB. Comparison to this model allows us to test how distinct is a point GW source vs an isotropic GWB using solely information in the correlations. 

The third is a cross-correlated CW model without $\psi$-dependence (\ref{eq:spike-pixel}), representing an unpolarized (or circularly polarized) point source, also referred to by \citet{Schult2025} as a \emph{spike-pixel} model for GW anisotropy. Comparison to this model tests how well the $\psi$ and $\iota$-dependent modulations can be resolved solely using correlations. Note that we injected a linearly polarized GW in the simulated dataset in order to maximize the differences between the spike-pixel and general BHB models.

\begin{figure*}
    \centering
    \includegraphics[width=0.9\linewidth]{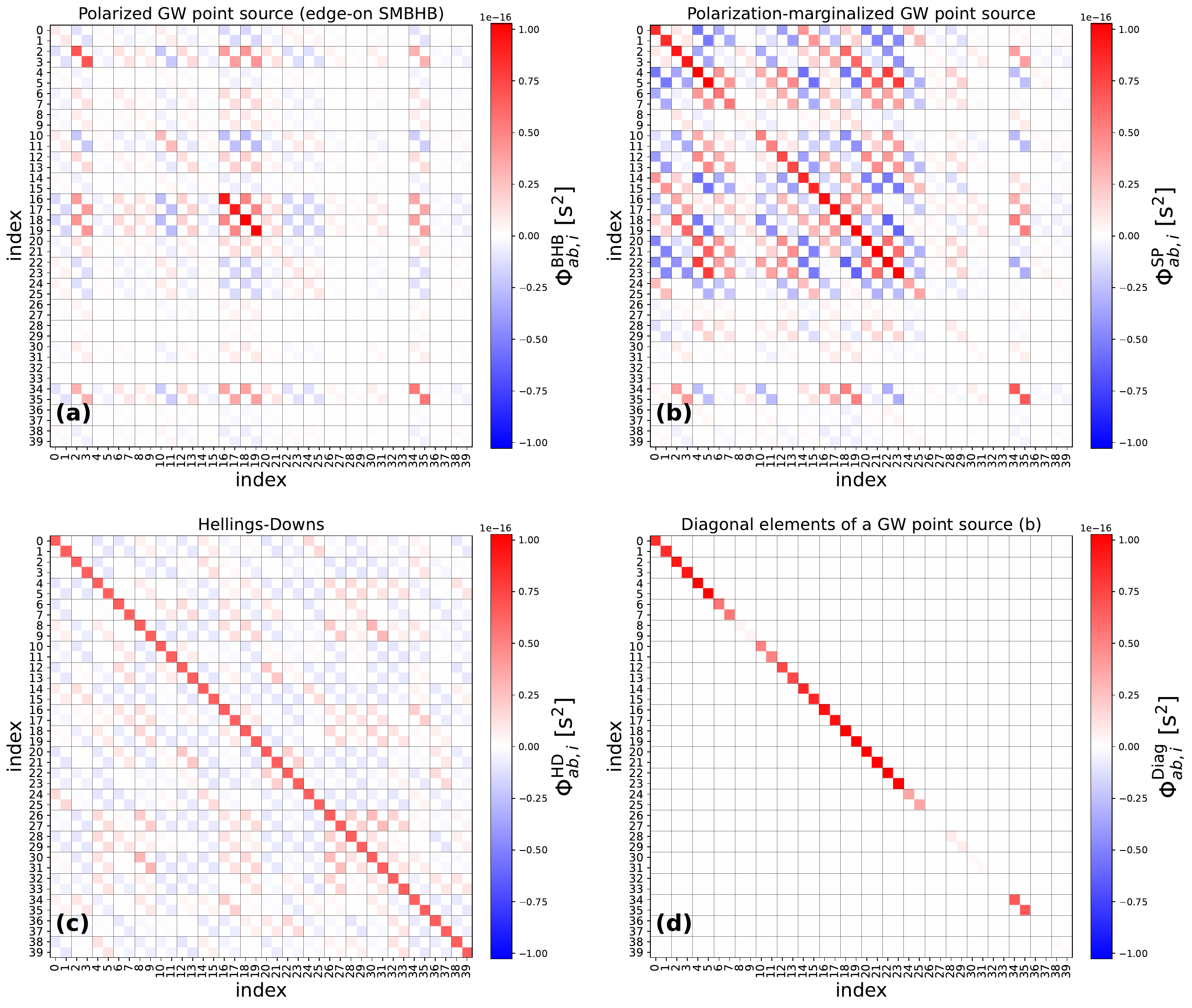}
    \caption{Sub-block of the CW covariance matrix $\bm{\Phi}$ [Eq.~\eqref{eq:Phi_matrix_CW}] for 4 different choices of ORF, $\Gamma_{ab}$. These 4 ORFs correspond to (a) an elliptically polarized GW from a SMBHB (BHB; the $\Upsilon_{ab}$ fingerprint derived here) [Eq.~\eqref{eq:gamma_psi}], (b) an unpolarized point GW source [Eq.~\eqref{eq:spike-pixel}], (c) a Hellings \& Downs correlated model [Eq.~\eqref{eq:HD-curve}], and (d) an uncorrelated CW-like model [Eq.~\eqref{eq:Gamma-auto}]. The total matrix is $2N_{\rm psr} \times 2N_{\rm psr}$, with the CW represented by two Fourier coefficients in each pulsar; these sub-blocks correspond to the first 20 pulsars of our simulated dataset. Matrices are constructed using the injected CW parameters (Table~\ref{tab:simulation}) and visually partitioned into $N_{\rm psr} \times N_{\rm psr}$ blocks to distinguish each set of interpulsar correlations.}
    \label{fig:Phi}
\end{figure*}

We gain some insights into these different modeling choices by inspecting the covariances directly. Fig.~\ref{fig:Phi} visualizes four different versions of the CW $\bm{\Phi}$ matrix corresponding to the different ORFs discussed above. The overall amplitude of each matrix is set by $h_0^2/(4\pi^2 f_{\rm gw}^2)$, while the relative amplitude between matrix elements is governed solely by $\Gamma^{\rm CW}_{ab}$.

The $\bm{\Phi}$ matrix for our polarized SMBHB source, including $\psi$ and $\iota$-dependence, $\bm{\Phi}^{\rm BHB}$, is shown in Fig.~\ref{fig:Phi}, panel (a). This features strong amplitude modulations along the matrix diagonal, based on the angular sky distance between the GW source and each pulsar, with distinct correlation fingerprints on the off-diagonals. We note note the oscillatory $\psi$ modulations in Eq.~\eqref{eq:gamma_psi}, which suppress GW power for many pulsars alongside the sky location parameters, are highly pronounced here, since we have assumed an edge-on orientation for the binary in this particular simulation. We note similar features using the spike-pixel CW model of panel (b), $\bm{\Phi}^{\rm SP}$, but the lack of $\psi$-modulations results in additional power along the diagonal and off-diagonal elements in comparison to $\bm{\Phi}^{\rm BHB}$. We further note that $\bm{\Phi}^{\rm SP}$ is applicable to a circularly polarized CW source, and since the featured simulation is for an edge-on binary, the most general elliptically polarized CW may be viewed as a linear combination of the covariance matrices $\bm{\Phi}^{\rm SP}$ and $\bm{\Phi}^{\rm BHB}$. The HD-correlated model in panel (c), $\bm{\Phi}^{\rm HD}$, also includes cross-correlations, but the cross-correlations are inaccurate for many pairs, and the diagonal elements are improperly ascribed equal amplitudes based on an isotropic sky-distribution of GW power. 

Finally, the diagonal model in panel (d), $\bm{\Phi}^{\rm Diag}$, neglects cross-correlations entirely, featuring only the diagonal elements of $\bm{\Phi}^{\rm SP}$ from panel (b). 

We next analyze the idealized datasets using the four different CW models corresponding to the $\bm{\Phi}$ matrices in Figure~\ref{fig:Phi}. We use \texttt{enterprise} to construct the likelihood and adopt uniform priors on six  parameters,
\begin{align}
    \log_{10} h_0 &\sim \mathcal{U}(-18,-10), \\
    \log_{10} f_{\rm gw} &\sim \mathcal{U}(-9,-8), \\
    \cos\theta_{\rm gw} &\sim \mathcal{U}(-1,1), \\
    \phi_{\rm gw} &\sim \mathcal{U}(0,2\pi), \\
    \cos\iota &\sim \mathcal{U}(-1,1), \\
    \psi &\sim \mathcal{U}(0,\pi/2).
\end{align}
Here the strain amplitude $h_0$ is dimensionless and the GW frequency $f_{\rm gw}$ is expressed in Hz, so that the prior $\log_{10} f_{\rm gw}\in[-9,-8]$ corresponds to $1$--$10\,$nHz. All six parameters are used only for the most general $\bm{\Phi}^{\rm BHB}$; models $\bm{\Phi}^{\rm Diag}$ and $\bm{\Phi}^{\rm SP}$ sample only in $(h_0, f_{\rm gw}, \cos\theta_{\rm gw}, \phi_{\rm gw})$, while model $\bm{\Phi}^{\rm HD}$ samples only in $(h_0, f_{\rm gw})$. To match the convention in \texttt{enterprise}, $\theta_{\rm gw}$ and $\phi_{\rm gw}$ point (in the cosmic rest frame) towards the GW source rather than the direction of GW propagation. Note the form of Eq.~\ref{eq:gamma_psi} used to construct $\bm{\Phi}^{\rm BHB}$ requires sampling only in the quadrant $\psi \in [0, \pi/2]$, as opposed to the hemisphere $\psi \in [0, \pi]$ which is normally required when using the deterministic template model to account for the degeneracy between $\psi$ and a constant phase offset. Eq.~\ref{eq:gamma_psi} similarly implies one may sample in $\cos^2\iota \in [0,1]$ as opposed to $\cos\iota \in [-1,1]$, but we note this requires use of an inverse square root prior for $\cos^2\iota$ to match the uniform prior in $\cos\iota$ required for a random binary orientation. We use the \texttt{nautilus} importance nested sampling code to generate posterior samples and estimate model evidences \citep{nautilus}.

\begin{figure*}
    \centering
    \includegraphics[width=0.8\linewidth]{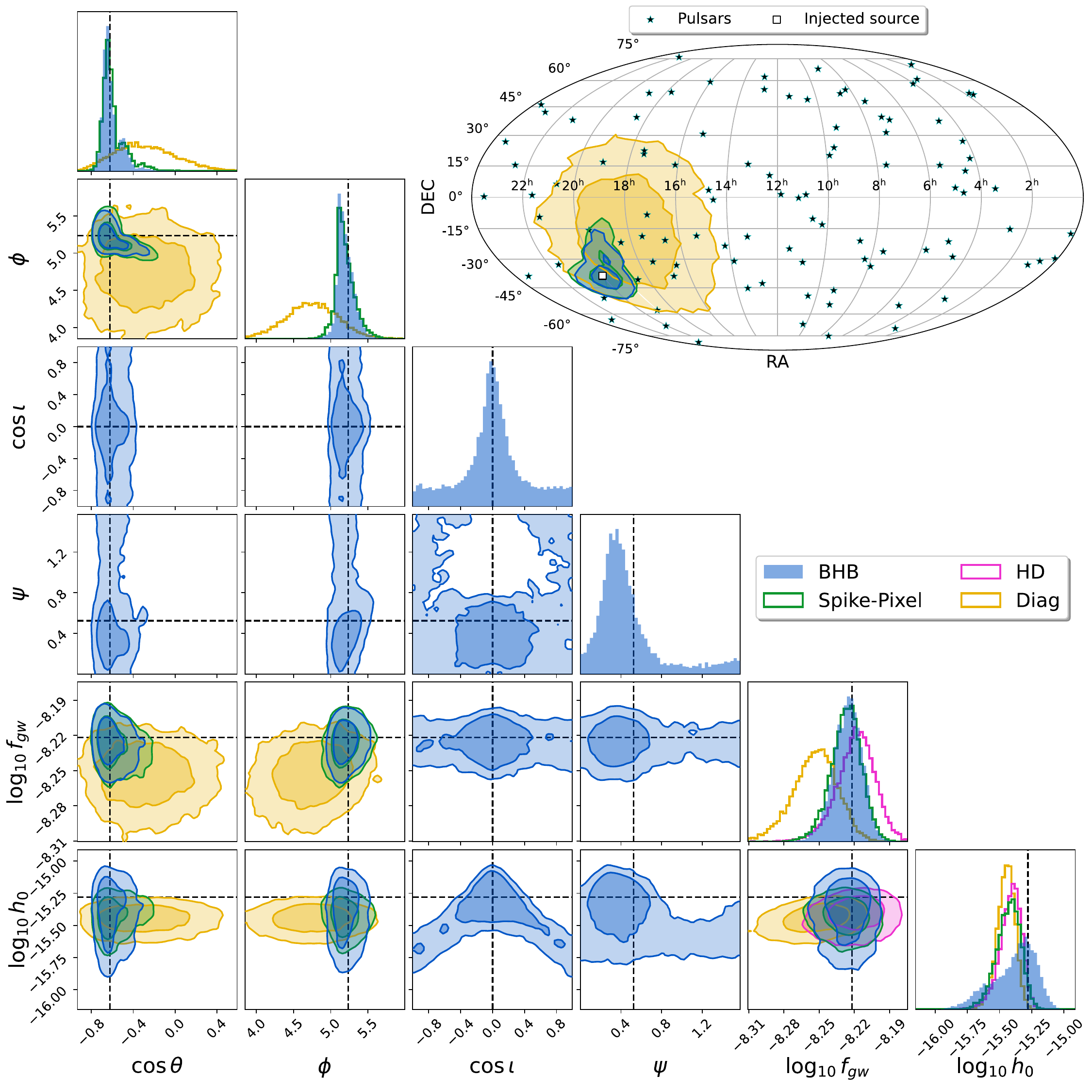}
    \caption{Posterior parameter distributions for cross-correlated CW models. Each model recovers the injected CW parameters (dashed black lines) in an idealized dataset where the injection uses a deterministic, evolving SMBHB template with ${\rm S/N}=30$ (Table~\ref{tab:simulation}). The physically-accurate cross-correlated CW models, spike-pixel (green; \citep{Schult2025}) and BHB (blue), yield more informative posteriors than the misspecified auto-correlated (yellow) and HD-correlated (pink; \citep{anholm2009}) models. Nonetheless, both the auto-correlated and HD-correlated models do recover a signal at nearly the correct frequency and strain amplitude. The inset skymap on the upper right shows that the cross-correlations of the spike-pixel and BHB models allows more precise and accurate source localization versus the purely auto-correlated model ($11\times$ smaller localization volume). By including $\psi$-dependence in the cross-correlations, the BHB model returns the most informative posteriors, recovering the edge-on binary orientation $\cos\iota = 0$. The BHB model also returns the most accurate $h_0$ posterior by resolving the impact of inclination angle $\iota$ on the signal amplitude. Nonetheless, frequency recovery and sky localization of the BHB model are mildly preferred to the spike-pixel model.}
    \label{fig:corner}
\end{figure*}

Using the highest S/N simulation, Table~\ref{tab:results} summarizes the detection and sky localization statistics for all four models, while Fig.~\ref{fig:corner} shows the resulting posteriors, with an inset showing the sky location posteriors projected onto a skymap with simulated pulsar locations. The most detailed CW model using $\bm{\Phi}^{\rm BHB}$ is able to correctly recover all six injected parameters, including $\cos\iota$ and $\psi$. We note the posteriors on $\cos\iota$ and $\psi$ are still not well constrained, highlighting the difficulty in recovering precise binary orientations using PTAs. Nonetheless, the $\bm{\Phi}^{\rm BHB}$-matrix model is able to recover the most accurate $h_0$ posterior out of the four models by decoupling the impact of inclination angle on the signal amplitude. The $\psi$-marginalized CW model, also known as the spike-pixel anisotropy model from \citep{Schult2025}, performs competitively well in source sky localization and frequency recovery as the $\psi$-dependent BHB model, despite not including the precise effects of binary orientation.

Comparison with the purely auto-correlated model given by $\bm{\Phi}^{\rm Diag}$ shows that valuable information is stored in the cross-correlations of the CW. Namely, the full cross–correlated models yield less error on the $\phi_{\rm gw}$ and $\cos\theta_{\rm gw}$ parameters, corresponding to a factor of $\sim11$ improvement in sky localization, as well as more accurate GW frequency recovery. Meanwhile, the HD-correlated model yields no sky information as the model assumes an isotropic distribution of GW power. Despite this misspecification, it can still be used to recover a CW signal with the correct GW frequency, i.e., it is possible to misclassify the signal as an isotropic GWB.

\begin{figure}
    \centering
    \includegraphics[width=\linewidth]{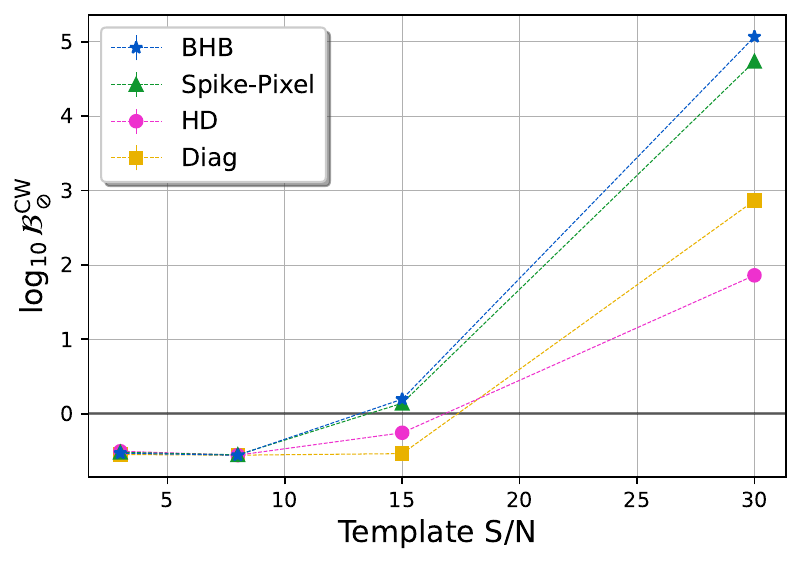}
    \caption{Bayes Factors for each of the four CW models compared against pulsar noise only, as a function of the injected dataset S/N, using our four idealized simulations with 100 pulsars (Table~\ref{tab:simulation}). We find that our cross-correlated CW model using the $\Upsilon_{ab}(\hat{\bm{\Omega}},\iota,\psi)$ fingerprint (blue; BHB) consistently recovers the highest Bayes Factor, followed closely by the spike-pixel model from \citet{Schult2025} for a face-on binary (green). Using the highest template ${\rm S/N}=30$ dataset, the misspecified auto-correlated (yellow; Diag) and HD-correlated (pink) models are also favored with $\mathcal{B}^{\rm CW}_\oslash > 1$. However, the correct model BHB is still favored with $\mathcal{B}^{\rm BHB}_{\rm Diag} = 159.13\pm 0.02$ and $\mathcal{B}^{\rm BHB}_{\rm HD} = 1611.37\pm 0.02$. While no model can detect the CW when the template ${\rm S/N} < 10$, use of cross-correlations can also begin to detect the CW with $\mathcal{B}^{\rm CW}_\oslash > 1$ in the $\rm{S/N} = 15$ case, while the auto-correlated and HD-correlated models cannot. These results demonstrate proof of concept that cross-correlations can be used for robust detection of CWs produced by a SMBHB and distinguished from auto-correlated noise as well as HD correlations from an isotropic GWB.}
    \label{fig:Bayes_facs}
\end{figure}

In order to further distinguish between models, we turn to the model evidences. We compute these as the integral of the likelihood times prior over the parameter volume using nested sampling, and then divide by the evidence of the data with noise alone, computed as $\mathcal{Z}_\oslash = \mathcal{L}(\bm{\delta t}|h_0=0)$, in order to determine a Bayes factor, which we denote $\mathcal{B}$. Fig.~\ref{fig:Bayes_facs} shows the resulting Bayes Factors for each model as a function of the S/N of the injected CW. Across the datasets with different S/N injections, we find the highest evidence in favor of our cross-correlated model for a general SMBHB orientation, as expected since this form most closely follows from the deterministic template model used for the injection. The power of our cross-correlated implementation, given by Eq.~\eqref{eq:Phi_matrix_CW}, lies in the ability to compare against other correlated model hypotheses, notably the auto-correlated (Diag) and HD-correlated models given by Eqs.~(\ref{eq:Gamma-auto}, \ref{eq:HD-curve}). In the intermediate ${\rm S/N}=15$ dataset, the cross-correlated model BHB can begin to detect the CW with $\mathcal{B} > 1$, while the HD and auto-correlated models cannot. In the high ${\rm S/N}=30$ dataset, both the Diag and HD models do detect the CW (as seen in Fig.~\ref{fig:corner}), however the evidence for the physically accurate BHB model is substantially higher, with $\mathcal{B}^{\rm BHB}_{\rm Diag} = 159.13\pm 0.02$ and $\mathcal{B}^{\rm BHB}_{\rm HD} = 1611.37\pm 0.02$. 

\begin{table}
    \centering
    \begin{tabular}{ c | c | c | c | c | c | c @{}}
        \hline\hline Model $\mathcal{M}$ & BHB & SP & HD & Diag & None & Ref.  \\
        \hline Bayes Factor $\mathcal{B}^{\rm BHB}_{\mathcal{M}}$ & 1 & 2.14 & 1611 & 159 & $10^5$ & Fig.~\ref{fig:Bayes_facs} \\
        \hline Localization $\Delta\Omega$ [${\rm deg}^2$] & 660 & 861 & -- & 7072 & -- & Fig.~\ref{fig:corner} \\
        \hline\hline
    \end{tabular}
    \caption{Statistics for the four different cross-correlated CW models, using the ${\rm S/N} = 30$ CW injection from Table~\ref{tab:simulation}. The first row gives the signal model $\mathcal{M}$ (as distinguished by the choice of ORF, with ``None'' for no CW), the second gives the Bayes Factor for the BHB-induced CW over the model $\mathcal{M}$, and the third gives the 95\% sky localization volume obtained using the model $\mathcal{M}$.}
    \label{tab:results}
\end{table}

This proof-of-concept result shows firstly that cross-correlation information can be leveraged to effectively distinguish the contributions of a CW from that of an isotropic GWB. Secondly, cross–correlations can yield a substantial Bayes factor in favor of a coherent CW model, even when the autocorrelations by themselves can be used to detect the CW. Although the Bayes factor $\mathcal{B}^{\rm BHB}_{\rm Diag}$ is smaller than $\mathcal{B}^{\rm BHB}_\oslash$ for the highest S/N dataset, the former statistic directly quantifies the additional information carried by the interpulsar fingerprints and will be less susceptible to biases from misspecified pulsar noise models~\cite{Goncharov2025, Larsen2026-noise, Agarwal2026-gwb-noise}.

While including $\iota$ and $\psi$-modulations to the ORF are in principle required for the most physically accurate ORFs, Fig.~\ref{fig:Bayes_facs} confirms that a model where $\psi$ and $\iota$ are marginalized over, equivalent to the spike-pixel anisotropy model from \citet{Schult2025}, is not strongly disfavored over the BHB model. Indeed, we find that $\mathcal{B}^{\rm BHB}_{\rm SP} = 2.14\pm 0.02$ in the highest ${\rm S/N}=30$ dataset. This again reflects the known challenges in precisely constraining binary orientations in PTA datasets \citep{SesanaVecchio2010_CW, petrov2025}, and is in line with claims from \citet{Schult2025} that accounting for GW polarization only impacts parameter recovery at high CW S/N. %

\section{Discussion}
\label{sec:discussion}
We have derived the direction-dependent ORF $\Upsilon_{ab}(\hat{\bm{\Omega}},\iota,\psi)$ for a single circular SMBHB in a PTA. This function acts as the deterministic analogue to the HD curve: just as the HD curve encodes the correlation signature of an isotropic background, $\Upsilon_{ab}$ encodes the unique spatial fingerprint of a single resolved binary. Working in the computational frame yields the compact analytic expression in Eq.~\eqref{eq:upsilon-final-psi}, or Eq.~\eqref{eq:upsilon-final} when marginalizing over binary orientation, providing a direct geometric expression for the spatial correlation pattern of an individual nanoHertz source. These expressions complement previous descriptions of the ORF of a point GW source, e.g. \citep{anholm2009, cornishsesana2013, Schult2025}, by highlighting the exact role of interpulsar separation angle $\zeta$, alongside binary sky location and orientation parameters, in shaping the ORF for a given pulsar pair.

This geometric framework unifies features observed in recent simulations, where bright binaries generate spectral variance~\cite{Becsy2022}, directional hotspots~\cite{Schult2025}, and frequency-dependent anisotropy~\cite{Gersbach+2025_PFOS}. As shown by Cornish and Sesana~\cite{cornishsesana2013} and Allen~\cite{Allen2023}, the HD curve is recovered only in the stochastic limit of averaging over many source locations. Consequently, deviations from the HD form in realistic datasets should be understood as incomplete averaging over these discrete $\Upsilon_{ab}$ patterns.

Our simulations in Sec.~\ref{sec:injections} demonstrate that this fingerprint is a vital practical tool for source identification. In standard spectral searches, a single bright CW source can easily masquerade as a stochastic background if only the power spectrum is modeled. However, by using the single source ORF, $\Upsilon_{ab}$, we successfully break this degeneracy, finding $\mathcal{B} \approx 1611$ in our highest S/N simulation. The data further favors the single-source model over a spatially uncorrelated model with a Bayes factor of $\mathcal{B} \approx 160$ in our highest S/N simulation, crucially demonstrating how CW cross-correlations may distinguish genuine CWs from noise artifacts, which has been a highly non-trivial problem to diagnose in historical CW analyses e.g., \citep{Aggarwal_2019, falxa_searching_2023, ng15-cw, ng15-targeted}. Furthermore, while auto-correlations capture a substantial portion of the signal power ($\sim 66\%$ in our injection), 
it is the cross-correlations that map the distinct spatial correlation pattern unique to that specific binary. Unlike the universal HD curve, this geometric fingerprint depends explicitly on the source location, providing the leverage required to break the degeneracy with the background.

A detailed comparison between fingerprint-based cross-correlation searches and fully coherent matched-filter CW pipelines is presented in dedicated follow-up work, and is also being explored in simulation campaigns such as \cite{Schult2025}. Here we clarify the trade-off between sensitivity and robustness. Coherent statistics that model the full Earth-plus-pulsar response offer higher theoretical sensitivity because they extract signal information from both ends of the line of sight. However, part of this gain is contingent on having pulsar distances precise to within a gravitational wavelength ($\delta L \ll \lambda_{\rm gw}$). Without such precision, the pulsar-term phases are unconstrained. Explicitly modeling them requires introducing $N_{\rm psr}$ additional nuisance parameters, which in turn increases the dimensionality of the parameter space and incurs a statistical penalty (or Occam factor) that reduces the significance of a detection. While recent work suggests this penalty may not always be prohibitive~\cite{Schult2025, petrov2025}, the robustness of a method that does not rely on these poorly constrained parameters remains a distinct advantage for initial detection. In the current regime, the cross-correlation statistic $\Upsilon_{ab}$ provides a pragmatic alternative. By treating the pulsar term as a nuisance parameter, $\Upsilon_{ab}$ formally discards the pulsar term, but gains a robust geometric factor that relies strictly on interpulsar correlations. Until overall pulsar distance measurements improve, this geometric robustness is incredibly valuable for detection. Where pulsar terms become the most useful given present pulsar distance constraints is to account for SMBHB frequency evolution, relevant for high frequency and chirp mass binaries. For future analyses of such binaries, a cross-correlated model for the CW must allow an explicit decoupling of the Earth and pulsar terms in the CW spectral design matrix to properly account for frequency evolution. 

Although our derivations assume circular binaries, the geometric structure is robust to moderate eccentricity. For typical PTA sources, the quadrupole ($n=2$) harmonic dominates, meaning the cross correlations remain governed by the circular fingerprint $\Upsilon_{ab}$~\cite{enoki_effect_2007, TaylorHuertaGair2016}. The framework also naturally extends to testing General Relativity: replacing the GR tensor polarization tensors with scalar or vector modes yields modified fingerprints with distinct angular structure~\cite{Eardley1973a, LeeJenetPrice2008, ChamberlinSiemens2012, Qin2021, ng15-altpol}, offering a clean probe of non-tensorial polarization content. We develop this direction in a companion paper~\citep{Zheng2026}, extending the single-source fingerprint formalism to tests of general relativity.

Finally, we look toward the era of high-precision astrometry. As summarized in Appendix~\ref{sec:pulsar-terms-CW}, accurate pulsar distances will eventually allow us to move beyond Earth-term fingerprints and exploit the pulsar terms coherently. This opens the door to measuring chirp masses and source distances via parallax, effectively turning the array into a galactic-scale interferometer~\cite{corbincornish2010, Mingarelli2012}. This transition will likely follow a hybrid path: using the robust Earth-term fingerprints $\Upsilon_{ab}$ to identify and localize candidates today, and adding coherent pulsar-term modeling for high-precision pulsars as data quality permits. Multimessenger constraints from the \textit{Fermi} PTA can further support this by identifying targeted millisecond pulsars near candidate GW sources~\cite{Kerr, fermi-pta}.

\section{Conclusions}
\label{sec:conclusion}
These results show that each SMBHB contributes a stable spatial correlation pattern, and that many features observed in simulations arise from superpositions of a small number of such patterns. When many binaries contribute with random orientations, the aggregate tends toward the HD curve. When only one or a few dominate, the sky retains the structure of the individual fingerprints derived here.

This framework naturally integrates into the evolving timeline of PTA discovery. As the community solidifies the evidence of the isotropic background via the HD correlation, the observational frontier will shift toward resolving the individual SMBHBs that stand out against this stochastic signal. The single source ORF, $\Upsilon_{ab}$, derived here provide the necessary tool for this regime, enabling the identification and separation of bright deterministic signals even where the background is strong. This step is essential for the ultimate goal of mapping the nanoHertz sky: once these distinct sources are resolved, the remaining anisotropy imprinted by the cosmic large-scale structure~\cite{Semenzato_2026, Allen2024} can be characterized using the spherical harmonic formalism established in~\cite{m13}. Together, these methods form a unified pipeline, leading us from the initial detection of the background to a complete astrophysical map of the cosmic SMBHB population.

\acknowledgments

\noindent The authors are grateful to B. Becsy, S. Taylor, L. Schult for comments on an early draft, and V. Ozolins for useful discussions. C.\ M.\ F.\ M.\ is grateful to the late Ahmed H. Zewail, who inspired this work with his 2016 lecture at Caltech. C.\ M.\ F.\ M.\ was supported in part by the National Science Foundation under Grants No.\ NSF PHY-1748958, NASA LPS 80NSSC24K0440, and NSF PHY-2309135 to the Kavli Institute for Theoretical Physics (KITP). C.M.F.M. thanks the Center for Computational Astrophysics (CCA) of the Flatiron Institute for support. The Flatiron Institute is supported by the Simons Foundation. E.E. was funded in part by a NASA-CT Space Grant, PTE Federal Award No: 80NSSC25M7127.

\section*{Software}
This work made use of \texttt{enterprise}~\citep{ENTERPRISE}, \href{https://github.com/bencebecsy/pta_replicator}{\texttt{pta\_replicator}} (created by Bence B{\'e}csy, Jeff Hazboun, and Aaron Johnson, with code adapted from Michele Vallisneri's \href{https://github.com/vallis/libstempo/tree/master}{\texttt{libstempo}}), and \texttt{nautilus}~\citep{nautilus}, together with the scientific Python ecosystem: \texttt{NumPy}~\citep{Numpy}, \texttt{SciPy}~\citep{Scipy}, \texttt{Matplotlib}~\citep{matplotlib}, and \texttt{Astropy}~\citep{Astropy}. We also acknowledge the use of Anthropic's Claude Code.

\appendix
\section{Antenna patterns }
\label{appendix:beam}

This appendix collects the expressions needed to reproduce the antenna
patterns and the rotation of an arbitrary PTA geometry into the
computational frame used in the main text. One can rotate the pulsars into their inherent source frame using the methods developed in \citet{m13}.

With pulsar $a$ placed on the $+\hat{z}$ axis and pulsar $b$ placed in
the $x$–$z$ plane at separation $\zeta$, the GW propagation direction
is parameterized by
\[
\hat{\bm{\Omega}}=(\sin\theta\cos\phi,\,
\sin\theta\sin\phi,\,
\cos\theta).
\]
We note that the transformation of the coordinates $\theta, \phi$ from pointing to the source to the GW propagation direction involves a parity transformation on the source coordinates. The transverse orthonormal basis vectors are
\begin{align}
    \hat{\bm{m}} &= (\sin\phi,-\cos\phi,0), \\
    \hat{\bm{n}} &=(\cos\theta\cos\phi,\,
    \cos\theta\sin\phi,\,
    -\sin\theta).
\end{align}
The polarization tensors are
\[
e^+_{ij}=\hat{m}_i\hat{m}_j-\hat{n}_i\hat{n}_j,\qquad
e^\times_{ij}=\hat{m}_i\hat{n}_j+\hat{n}_i\hat{m}_j.
\]

Contracting with $\hat{\bm{p}}_a=(0,0,1)$ gives
\begin{align}
F_a^+ &= -\frac{1}{2}\frac{\sin^2\theta}{1+\cos\theta} = -\frac{1}{2}(1-\cos\theta),\\
F_a^\times &= 0.
\end{align}
For pulsar $b=(\sin\zeta,0,\cos\zeta)$ we obtain
\begin{align}
F_b^+ &= \frac{1}{2}
\frac{(\hat{\bm{p}}_b\!\cdot\!\hat{\bm{m}})^2
-(\hat{\bm{p}}_b\!\cdot\!\hat{\bm{n}})^2}
{1+\hat{\bm{\Omega}}\!\cdot\!\hat{\bm{p}}_b} \\
&= -\frac{1}{2}\frac{(\cos\theta\sin\zeta\cos\phi - \sin\theta\cos\zeta)^2 - \sin^2\zeta\sin^2\phi}{1+\cos\theta\cos\zeta + \sin\theta\sin\zeta\cos\phi},\\
F_b^\times &=
\frac{(\hat{\bm{p}}_b\!\cdot\!\hat{\bm{m}})
(\hat{\bm{p}}_b\!\cdot\!\hat{\bm{n}})}
{1+\hat{\bm{\Omega}}\!\cdot\!\hat{\bm{p}}_b} \\
&= \frac{\sin\zeta\sin\phi(\cos\theta\sin\zeta\cos\phi - \sin\theta\cos\zeta)}{1+\cos\theta\cos\zeta + \sin\theta\sin\zeta\cos\phi},
\end{align}
with all intermediate contractions given explicitly above.
These expressions, when substituted into
$\Upsilon_{ab}(\theta,\phi,\zeta)=\alpha_a\alpha_b+\beta_a\beta_b$,
yield Eq.~\eqref{eq:upsilon-final}.

\section{Averaging \texorpdfstring{$\Upsilon_{ab}$}{Upsilon\_ab} over inclination and polarization angle}
\label{app:averaged_upsilon}

In this appendix we derive the single-source ORF averaged over the inclination angle $\iota$ and polarization angle $\psi$. Starting from Eq.~(\ref{eq:upsilon-final-psi}),
\begin{align}
\Upsilon_{ab}(\theta,\phi,\zeta,\iota,\psi) &= \tfrac{1}{2}\bigl[1 + \sin^4\!\iota\cos4\psi \nonumber\\
&\quad + \cos^4\!\iota + 6\cos^2\!\iota\bigr] F^+_a F^+_b \nonumber \\
&\quad + \tfrac{1}{2}\sin^4\!\iota\sin4\psi\; F^+_a F^\times_b\,,
\label{eq:upsilon_full}
\end{align}
where $F^+_a$, $F^+_b$, and $F^\times_b$ are the antenna pattern functions in the computational frame, and we have used $F^\times_a = 0$.

\subsection{Averaging over \texorpdfstring{$\psi$}{psi}}

Averaging uniformly over the polarization angle $\psi \in [0, 2\pi)$, we use
\begin{equation}
\langle \cos 4\psi \rangle_\psi = \langle \sin 4\psi \rangle_\psi = 0\,.
\end{equation}
The second term in Eq.~(\ref{eq:upsilon_full}) vanishes identically, and the $\sin^4\!\iota\cos4\psi$ contribution in the first term is eliminated, yielding
\begin{equation}
\langle \Upsilon_{ab} \rangle_\psi = \frac{1}{2}\left(1 + \cos^4\!\iota + 6\cos^2\!\iota\right) F^+_a F^+_b\,.
\label{eq:upsilon_psi_avg}
\end{equation}

\subsection{Averaging over \texorpdfstring{$\iota$}{iota}}

We average over the inclination assuming a uniform distribution in $\cos\iota$, i.e.\ $\cos\iota \sim \mathcal{U}(-1,1)$. Defining $x = \cos\iota$, the required moments are
\begin{equation}
\langle x^2 \rangle = \frac{1}{2}\int_{-1}^{1} x^2\, dx = \frac{1}{3}\,, \qquad \langle x^4 \rangle = \frac{1}{2}\int_{-1}^{1} x^4\, dx = \frac{1}{5}\,.
\end{equation}
Substituting into Eq.~(\ref{eq:upsilon_psi_avg}),
\begin{equation}
\left\langle 1 + \cos^4\!\iota + 6\cos^2\!\iota \right\rangle_\iota = 1 + \frac{1}{5} + 6 \times \frac{1}{3} = \frac{16}{5}\,,
\end{equation}
and thus
\begin{equation}
\boxed{\langle \Upsilon_{ab} \rangle_{\iota,\psi} = \frac{8}{5}\, F^+_a\, F^+_b\,.}
\label{eq:upsilon_averaged}
\end{equation}

\section{Pulsar terms in CW analyses}
\label{sec:pulsar-terms-CW}

In this section we give a brief and balanced overview of the impact of pulsar terms on CW searches, focusing on when they carry useful information and when they are best treated as a nuisance. We frame the discussion in terms of current PTAs, where pulsar distances are poorly known relative to a gravitational wavelength, and a future regime where distances are measured to better than a wavelength.

For a monochromatic circular binary, the timing residual of pulsar \(a\) can be written as
\begin{equation}
s_a(t) = s_a^{\rm E}(t) - s_a^{\rm P}(t) ,
\end{equation}
where the Earth term and pulsar term are
\begin{align}
s_a^{\rm E}(t) &= A_a \cos\!\left[\Phi(t)\right] ,\\
s_a^{\rm P}(t) &= A_a \cos\!\left[\Phi(t - \tau_a)\right] .
\end{align}
Here \(A_a\) and \(\Phi(t)\) encode the binary and antenna pattern parameters, and
\begin{equation}
\tau_a \equiv \frac{L_a}{c} \left(1 + \hat{\bm{\Omega}}\cdot\hat{\bm{p}}_a\right)
\end{equation}
is the geometric time delay between the Earth and pulsar terms for a source in direction \(\hat{\bm{\Omega}}\) and a pulsar at distance \(L_a\) along \(\hat{\bm{p}}_a\). The pulsar term samples the same binary at an earlier time \(t_{\rm pulsar} = t - \tau_a\), typically \(10^3\)–\(10^4\) yr in the past for PTA baselines.

In terms of the phase, one can write
\begin{equation}
k_a \equiv \omega L_a (1 + \hat{\bm{\Omega}}\cdot\hat{\bm{p}}_a) = 2\pi f L_a (1 + \hat{\bm{\Omega}}\cdot\hat{\bm{p}}_a) ,
\end{equation}
so that the pulsar term effectively carries an extra phase \(k_a\) relative to the Earth term. The key control parameter is then the distance uncertainty relative to the gravitational wavelength
\begin{equation}
\lambda_{\rm gw} = \frac{c}{f} .
\end{equation}
If the fractional distance error \(\delta L_a\) satisfies \(\delta L_a \gtrsim \lambda_{\rm gw}\), the induced uncertainty on \(k_a\) is of order \(2\pi\) and the pulsar phase is effectively unconstrained. If instead \(\delta L_a \ll \lambda_{\rm gw}\), then \(k_a\) is known to better than order unity and the pulsar term can be modeled coherently.

When pulsar distances are known to better than a wavelength, the pulsar terms carry several types of additional information beyond the Earth term.

First, they probe binary evolution on a much longer effective baseline. The phase difference between Earth and pulsar terms,
\begin{equation}
\Delta\Phi_a \equiv \Phi(t) - \Phi(t - \tau_a) ,
\end{equation}
is sensitive to the intrinsic frequency \(f\), its time derivative \(\dot{f}\), and higher derivatives. \citet{corbincornish2010} showed that including the pulsar term and treating the pulsar distance as a parameter can double the effective signal power and significantly improve recovery of intrinsic binary parameters, including the chirp mass, when the frequency evolution is measurable. \citet{Mingarelli2012} emphasized that the pulsar term provides a direct view of the same SMBHB thousands of years earlier in its inspiral. In principle, a coherent Earth plus pulsar analysis can then constrain mass ratio and spins through the imprint of higher order post-Newtonian and spin effects in the phase, provided that the signal is strong enough and the system is sufficiently close to merger.

Second, pulsar terms can sharpen sky localization and source geometry. \citet{Lee2011} carried out a Fisher matrix study of single-source parameter estimation and found that the pulsar term is essential for accurate sky localization in the high signal-to-noise regime. The time delays \(\tau_a \propto L_a (1 + \hat{\bm{\Omega}}\cdot\hat{\bm{p}}_a)\) and the way the pulsar term projects onto each pulsar effectively create additional baselines and angular structure, improving constraints on the sky position, inclination, and polarization angle once \(L_a\) is under control.

Third, there is a potential bright siren application. If one can measure \(f\) and \(\dot{f}\) from the combined Earth-and-pulsar phase evolution, then in GR these quantities are tied to the chirp mass and luminosity distance. \citet{Mingarelli2012} discussed how coherent Earth plus pulsar observations could, in principle, be used to infer both masses and spins of SMBHBs. 

Finally, the roles can be reversed. \citet{Lee2011} showed that a strong GW signal can be used to improve the determination of pulsar distances themselves, through timing parallax and the structure of the combined Earth and pulsar response. In that sense, CW sources can act as calibrators for the PTA.

However, the science gains above come with significant computational costs.
Including pulsar terms coherently requires treating the pulsar distances, or equivalently the phases \(k_a\), as parameters. \citet{corbincornish2010} explicitly included one distance parameter per pulsar in a Bayesian analysis and showed that the dimensionality and structure of the likelihood become substantially more complex. The parameter space is highly correlated, especially when the signal-to-noise ratio is modest, and efficient sampling requires careful proposals in the space of eigenparameters. These issues become more severe for large PTAs and extensive multi source searches.

Moreover, for realistic current PTAs most distances are not known to better than a wavelength at nanoHertz frequencies, or a few parsecs. In that regime the pulsar phases cannot be treated as known. One has to marginalize over them, either explicitly or implicitly through broad distance priors. This marginalization washes out the coherent angular structure of the pulsar term. What remains is additional variance and parameter volume, not a clean spatial pattern that can be exploited in the same way as the Earth term.

This is why several works treat the pulsar term as a noise term in the current regime. For stochastic backgrounds, it is standard to show that after averaging over many sources the pulsar term contributions drop out of the cross correlation, leaving the HD curve as the dominant angular dependence \citep{MM18}. 

It is important to note, however, that the pulsar term does not strictly vanish. As shown in \citet{MS14} and proven generally in \cite{MM18}, the pulsar term is always present in the response. For a single source where no sky averaging occurs, it manifests as a rapidly oscillating function of the pulsar distance rather than some small value.

Given these complications, it is natural to ask when one can safely ignore the pulsar term and work with Earth term only templates. \citet{Charisi2024} addressed this question for targeted CW searches and showed that an Earth term approximation, in which the pulsar term is dropped, provides constraints on the total mass and GW frequency that are very similar to those obtained with a full Earth-plus-pulsar treatment for non evolving binaries, while being more than two orders of magnitude faster. Their framework is designed for large scale, multimessenger searches for SMBHB candidates identified electromagnetically, where computational efficiency is critical.

Our fingerprint \(\Upsilon_{ab}(\hat{\bm{\Omega}})\) is defined in precisely this Earth term spirit. In the regime where \(\delta L_a \gtrsim \lambda_{\rm gw}\) for most pulsars, marginalization over distance uncertainties removes the coherent pulsar term structure from the expectation value of the cross correlation. The Earth term then provides the correct angular dependence, and the pulsar term acts mainly as a noise source that increases variance but does not define a new angular basis. This is consistent with the findings of  \citet{corbincornish2010}, \citet{Lee2011}, \citet{Mingarelli2012}, and \citet{Charisi2024} in the low evolution, low distance precision regime.

In a future PTA with precise pulsar distances for a significant fraction of the array, the balance will shift. At that point it will become attractive to generalize Earth term fingerprint analyses like \(\Upsilon_{ab}\) to include coherent pulsar term structure for those pulsars with \(\delta L_a \ll \lambda_{\rm gw}\). This hybrid strategy would mirror the regimes explored in the studies above: use the Earth term to drive detection and define the robust angular pattern, while selectively exploiting pulsar terms where distance and signal quality justify the added complexity.

In the end the pulsar term is both an opportunity and a challenge. In the current PTA regime it is safer to treat it as a nuisance and build geometric fingerprints from just the Earth term. In a future high precision array, it will become a powerful handle on binary evolution, sky localization, and possibly distance, but at the cost of substantially more complex statistical inference.

\section{Circular polarization in GR}
\label{sec:AppC-circ}

In GR the metric perturbation can be written in the linear basis as
\begin{equation}
h_{ij}(t,\hat\Omega)=h_+(t)\,e^+_{ij}(\hat\Omega)+h_\times(t)\,e^\times_{ij}(\hat\Omega).
\end{equation}
Switching to the circular basis $(R,L)$,
\begin{equation}
e^R_{ij}=\frac{1}{\sqrt{2}}\big(e^+_{ij}+i\,e^\times_{ij}\big), \qquad
e^L_{ij}=\frac{1}{\sqrt{2}}\big(e^+_{ij}-i\,e^\times_{ij}\big),
\end{equation}
with mode amplitudes
\begin{equation}
h_R=\frac{1}{\sqrt{2}}(h_+-i\,h_\times), \qquad
h_L=\frac{1}{\sqrt{2}}(h_++i\,h_\times),
\end{equation}
so that
\begin{equation}
h_{ij}=h_R\,e^R_{ij}+h_L\,e^L_{ij}.
\end{equation}

A purely right–hand circularly polarized wave has $h_L=0$ and $h_R(t)=h_0 e^{i\Phi(t)}$, which implies
\begin{equation}
h_+(t)=\frac{h_0}{\sqrt{2}}\cos\Phi(t), \qquad 
h_\times(t)=\frac{h_0}{\sqrt{2}}\sin\Phi(t),
\end{equation}
with the opposite phase for left–hand polarization.

The PTA antenna patterns transform in the same way:
\begin{equation}
F_a^R=\frac{1}{\sqrt{2}}\big(F_a^+ - i\,F_a^\times\big), \qquad
F_a^L=\frac{1}{\sqrt{2}}\big(F_a^+ + i\,F_a^\times\big),
\end{equation}
and the timing response of pulsar $a$ to a right–hand circularly polarized wave is simply  
\begin{equation}
s_a(t)=F_a^R(\hat\Omega)\,h_R(t)\qquad (h_L=0).
\end{equation}

This basis change does not alter the underlying geometry: the GR circularly polarized fingerprints are identical to the tensor ones up to an overall phase and amplitude rescaling.

For a single right–hand circularly polarized wave we have
\begin{equation}
s_a(t)=F_a^R(\hat\Omega)\,h_R(t), \qquad 
s_b(t)=F_b^R(\hat\Omega)\,h_R(t),
\end{equation}
with $F_a^R=(F_a^+ - iF_a^\times)/\sqrt{2}$ and $h_R(t)=h_0 e^{i\Phi(t)}$.  
The Earth–term cross correlation is then proportional to
\begin{equation}
\left\langle s_a(t)s_b(t)\right\rangle 
\propto 
\Re\!\left[F_a^R F_b^{R*}\,\left\langle h_R h_R^*\right\rangle\right].
\end{equation}
\begin{widetext}
    
Using the definition of $F_a^R$,
\begin{equation}
F_a^R F_b^{R*}
=\frac{1}{2}\left(F_a^+ - iF_a^\times\right)\left(F_b^+ + iF_b^\times\right)
=\frac{1}{2}\Big(F_a^+F_b^+ + F_a^\times F_b^\times + i\big[F_a^+F_b^\times - F_a^\times F_b^+\big]\Big),
\end{equation}

\end{widetext}
so that
\begin{equation}
\Re\!\left[F_a^R F_b^{R*}\right]
=\frac{1}{2}\left(F_a^+F_b^+ + F_a^\times F_b^\times\right).
\end{equation}
The circularly polarized fingerprint can therefore be written as
\begin{equation}
\Upsilon^{(R)}_{ab}(\hat\Omega)
\propto 
F_a^+(\hat\Omega)F_b^+(\hat\Omega)
+
F_a^\times(\hat\Omega)F_b^\times(\hat\Omega),
\label{eq:upsilon-circbasis}
\end{equation}
up to an overall factor $1/2$ that can be absorbed into the amplitude normalization.  
The same expression holds for a purely left–hand circularly polarized wave, with $F_a^L$ in place of $F_a^R$.

Thus the geometric dependence of the circularly polarized GR fingerprints is identical to that of the standard tensor case: switching from an unpolarized GW to a circularly polarized GW changes only the complex phase and overall amplitude, not the angular structure of $\Upsilon_{ab}(\hat{\bm{\Omega}})$.

Lastly, Eq.~\eqref{eq:upsilon-circbasis} and its linear polarization counterparts are derived assuming a fixed set of transverse polarization basis vectors in the computational frame, with $\hat{\bm{m}}$ lying along the $xy$-plane. In the cosmic rest frame, these basis vectors are tied to the binary's orientation via rotation of the basis vectors by some angle $\psi$ about $\hat{\bm{\Omega}}$. It's straightforward using the circular polarization basis to verify that Eq.~\eqref{eq:upsilon-circbasis}, the result for circular polarization is invariant to any arbitrary $\psi$ rotation (which must naturally be true, as Eq.~\eqref{eq:upsilon-circbasis} is also the result for an unpolarized GW). The circular polarization tensors may be rewritten,
\begin{align}
    e_{ij}^{(R/L)} &= (\hat{m}_i \pm i\hat{n}_i)(\hat{m}_j \pm i\hat{n}_j).
\end{align}
where we can see that the real and imaginary parts correspond to the typical plus and cross polarizations. Next, we perform a transformation that rotates the polarization vectors $\hat{\bm{m}} \to \hat{\bm{m}}'$, $\hat{\bm{n}} \to \hat{\bm{n}}'$ by an angle $\psi$. The transformation is defined
\begin{align}
    \begin{pmatrix}
        \hat{\bm{m}}' \\
        \hat{\bm{n}}'
    \end{pmatrix} = \begin{pmatrix}
        \cos\psi & -\sin\psi \\
        \sin\psi & \cos\psi
    \end{pmatrix}\begin{pmatrix}
        \hat{\bm{m}} \\
        \hat{\bm{n}}
    \end{pmatrix}.
\end{align}
Following this definition, the polarization tensors an antenna beam patterns transform as
\begin{align}
    (e_{ij}^{(R/L)})' &= e^{\pm2i\psi}e_{ij}^{(R/L)}, \\
    (F^{(R/L)})' &= e^{\pm2i\psi}F^{(R/L)}.
\end{align}
This yields the result that the ORF is invariant to $\psi$ rotations,
\begin{align}
    (\Upsilon^{(R/L)}_{ab})' &= \Re[({F^{(R/L)}_a}^*)'(F^{(R/L)}_b)']\nonumber \\
    &= \Re[e^{-2i\psi}({F^{(R/L)}_a}^*)e^{2i\psi}({F^{(R/L)}_b})]\nonumber \\
    &= \Upsilon^{(R/L)}_{ab}.
\end{align}
As such, for either a circularly polarized or a statistically unpolarized CW we are free to choose the polarization basis vectors in the computational frame. 

\bibliographystyle{apsrev4-2}
\bibliography{bib-clean}

\end{document}